\documentclass[aps,prx,reprint,superscriptaddress,nofootinbib,longbibliography,floatfix]{revtex4-2}

\usepackage{amsmath,amssymb,amsthm,mathtools}
\usepackage[T1]{fontenc}
\usepackage{lmodern}
\usepackage{graphicx}
\usepackage{booktabs}
\usepackage{xcolor}
\usepackage[normalem]{ulem}
\usepackage[hidelinks]{hyperref}
\makeatletter
\newcommand{\sharedcorresponding@number}{5}
\newcommand{\sharedcorrespondingthanks}{%
  \expandafter\sharedcorresponding@wrap\@author
}
\def\sharedcorresponding@wrap#1#2{%
  \def\@author{{#1}{\sharedcorresponding@note{#2}}}%
}
\let\sharedcorresponding@footnote\frontmatter@footnote
\def\frontmatter@footnote#1{%
  \in@{\sharedcorresponding@note}{#1}%
  \ifin@
    #1%
  \else
    \sharedcorresponding@footnote{#1}%
  \fi
}
\def\sharedcorresponding@note#1{%
  \sharedcorresponding@footnote{#1}%
  \textsuperscript{,\,}\frontmatter@footnotemark{\sharedcorresponding@number}%
}
\appdef\frontmatter@footnote@produce{%
  \present@FM@footnote{\sharedcorresponding@number}{%
    These authors contributed equally as corresponding authors.}%
}
\makeatother

\newtheorem{lemma}{Lemma}
\newtheorem{theorem}{Theorem}

\newtheorem*{corollary}{Corollary}
\theoremstyle{plain}

\theoremstyle{remark}

\newcommand{\Omax}{\Omega_{\max}}
\newcommand{\TR}{T_{R}}
\newcommand{\dd}{\mathrm{d}}
\newcommand{\ee}{\mathrm{e}}
\newcommand{\ii}{\mathrm{i}}
\newcommand{\bket}[1]{|#1\rangle}
\newcommand{\bbra}[1]{\langle #1|}

\newcommand{\xv}{\boldsymbol{x}}
\newcommand{\zv}{\boldsymbol{z}}

\begin{document}
\title{Certifying the Rydberg-exposure limit of a globally driven controlled-\texorpdfstring{$Z$}{Z} gate}
\author{Yusheng Zhao}
\thanks{These authors contributed equally to this work.}
\affiliation{QudeLeap Research, Shanghai 200030, China}
\affiliation{The Hong Kong University of Science and Technology (Guangzhou), Guangdong 511453, China}
\author{Han Wang}
\thanks{These authors contributed equally to this work.}
\affiliation{QudeLeap Research, Shanghai 200030, China}
\affiliation{The Hong Kong University of Science and Technology (Guangzhou), Guangdong 511453, China}
\author{Mingrui Jing}
\affiliation{QudeLeap Research, Shanghai 200030, China}
\affiliation{The Hong Kong University of Science and Technology (Guangzhou), Guangdong 511453, China}
\author{Zhongyi Ni}
\affiliation{QudeLeap Research, Shanghai 200030, China}
\affiliation{The Hong Kong University of Science and Technology (Guangzhou), Guangdong 511453, China}
\author{Hongshun Yao}
\affiliation{QudeLeap Research, Shanghai 200030, China}
\affiliation{The Hong Kong University of Science and Technology (Guangzhou), Guangdong 511453, China}
\author{Xin Wang}
\email[Contact author: ]{wangxinfelix@gmail.com}
\affiliation{The Hong Kong University of Science and Technology (Guangzhou), Guangdong 511453, China}
\author{Jie Wang}
\email[Contact author: ]{wangjie212@amss.ac.cn}
\sharedcorrespondingthanks
\affiliation{Academy of Mathematics and Systems Science, Chinese Academy of Sciences, Beijing 100190, China}
\author{Jin-Guo Liu}
\email[Contact author: ]{jinguoliu@hkust-gz.edu.cn}
\sharedcorrespondingthanks
\affiliation{The Hong Kong University of Science and Technology (Guangzhou), Guangdong 511453, China}
\date{\today}

\begin{abstract}
Rydberg excitation enables entangling gates in neutral-atom
quantum processors but also exposes them to spontaneous decay.  Whether this exposure can
be made arbitrarily small through pulse shaping and longer gate times has remained
unresolved. We show that an exact controlled-$Z$ gate driven by a single global
laser has an irreducible Rydberg-exposure cost in the perfect-blockade regime.
The product of
the maximum Rabi frequency $\Omega_{\max}$ and the average time $T_R$ spent in the
Rydberg state is bounded below by $2.9312$, independent of pulse shape and duration. We derive this bound using a sum-of-squares certificate, strengthened by constraints imposed by the shared drive.
Using interval arithmetic, we also certify the existence of an exact gate pulse
with $\Omax\TR<2.9359$.
The unrounded bounds differ by less than $0.16\%$.
The bound identifies a fundamental cost of shared-drive entangling
dynamics and establishes a benchmark against which future gate-control strategies
can be assessed.
\end{abstract}
\maketitle

\section{Introduction}
\label{sec:intro}

Fault-tolerant quantum computation requires low-error physical operations
and repeated quantum error correction. Reconfigurable neutral-atom arrays
are a leading platform for this goal, combining high-fidelity
Rydberg-mediated controlled-$Z$ (CZ) gates with programmable connectivity and
logical-qubit operations~\cite{Saffman10,Levine19,Evered23,Bluvstein26}.
Atom loss is a major source of error: in a recent CZ-gate experiment,
it accounted for approximately $60\%$ of the total gate
error~\cite{Evered26}. Unlike a Pauli error, loss removes the physical
carrier of the qubit. It can disrupt subsequent gates and stabilizer
measurements, complicating decoding, while sustained operation requires
replacing missing atoms. The architecture demonstrated by Bluvstein
\emph{et al.} uses an atom reservoir and mid-circuit replacement for this
purpose~\cite{Bluvstein26,Baranes26}.

Reducing loss during gate operations can complement the replacement of lost atoms.
Rydberg-state decay during CZ gates is one of the main gate-error
mechanisms and contributes to atom loss~\cite{Evered23,Evered26}.
The same excitation that enables
blockade-mediated entanglement exposes the atoms to finite radiative and
blackbody-limited lifetimes~\cite{Jaksch00,Lukin01,Beterov09}.
The relevant cost is the average Rydberg exposure $\TR$: the
physical-time integrated Rydberg population, averaged equally over the
four computational-basis inputs. Assuming that each decay event at
single-atom rate $\Gamma$ irreversibly removes population from the
computational subspace, the input-averaged decay-loss probability is
$P_{\mathrm{loss}}\approx\Gamma\TR$ to the first
order~\cite{JP22,Locher25,Pagano22,Jandura23}.
Reducing this contribution therefore requires reducing exposure.

Can a carefully designed CZ pulse make this exposure arbitrarily small
at fixed available laser strength? A longer gate might keep the Rydberg
populations low enough to reduce their time integral. For an isolated
two-level atom, this possibility is realized: an arbitrary ground-state
phase can be acquired with exposure tending to zero through a sequence
of increasingly detuned, longer pulses (Appendix~\ref{sec:nogo}).
However, a shared laser field, used in global-drive CZ
gates~\cite{Levine19,Evered23}, ties the single- and two-atom excitation
sectors to the same control with coupling ratio $\sqrt{2}$.
We seek the infimum $C_*:=\inf \Omax\TR$ over all exact,
finite-duration, globally driven CZ gates in the perfect-blockade regime,
with Rabi-frequency envelope $\Omega(t)$ bounded by a fixed $\Omax$.
The dimensionless exposure $C=\Omax\TR$ separates pulse design from drive strength.

Alongside analytic pulse shaping and detuning modulation~\cite{Theis16,Shi17,Shi19},
numerical optimal control methods~\cite{Khaneja05,Caneva11,Koch22} have found
low-exposure waveforms.
Jandura and Pupillo~\cite{JP22} reported $C\approx2.957$ for a
time-optimal CZ pulse and $C\approx2.947$ after optimizing exposure.
Fromonteil \emph{et al.}~\cite{Fromonteil24} subsequently recovered the globally
time-optimal CZ trajectory using Hamilton--Jacobi--Bellman equations, which
minimize gate duration rather than exposure.
More recently, Locher \emph{et al.}~\cite{Locher25} reduced the exposure to
$C\approx2.936$.
These exposure values provide numerical benchmarks but do not
establish a lower bound valid for all pulse shapes and durations.
For maximally entangled-state preparation, Doultsinos \emph{et al.}~\cite{DDP25}
derived a decay-error bound proportional to $\Gamma/B$ at finite interaction
strength $B$. This bound vanishes in the perfect-blockade limit at fixed laser
strength and therefore does not determine the exposure floor for exact global
CZ synthesis.

Polynomial optimization and convex relaxations provide several
approaches to quantum control, including conservation-law-based semidefinite
bounds~\cite{Zhang21}, global optimization of polynomial approximations to
controlled propagators~\cite{Bondar25}, and moment--sum-of-squares performance
bounds for open-loop and feedback control~\cite{HoltorfThesis24,Holtorf24}.
Naceur and Balada Gaggioli~\cite{Naceur26} developed certificates for
reachability and minimum control times with a fixed upper bound on the control
duration. We seek a fundamental lower bound on Rydberg exposure that remains
valid for arbitrarily long gates. This requires a formulation without a
prescribed maximum gate duration.

We show that a shared drive imposes a nonzero Rydberg-exposure floor
on exact CZ gates, even when gate duration is unrestricted. We establish
certified two-sided bounds with a relative gap below $0.16\%$.
Rounded outward, these bounds read
\begin{equation}
\boxed{2.9312\le C_*<2.9359.}
\label{eq:headline}
\end{equation}
The exact endpoints are stated in Theorem~\ref{thm:slack_strengthened} and
Eq.~\eqref{eq:upper_bound}; the relative gap uses those unrounded values.

\begin{figure*}[t]
\centering
\includegraphics[width=0.8\textwidth,trim=0mm 4.8mm 0mm 0.7mm,clip]{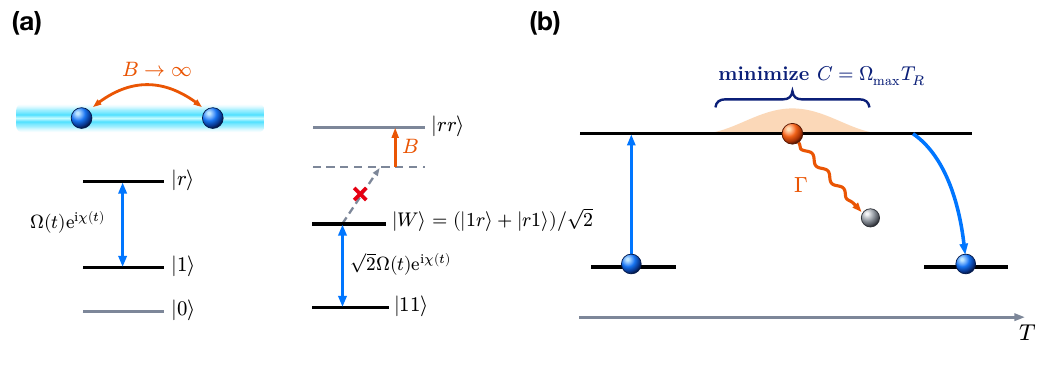}
\caption{{Shared-drive Rydberg gate and exposure to decay.
(a)~A shared laser field with envelope $0\le\Omega(t)\le\Omax$ and phase
$\chi(t)$ couples $\bket{1}\leftrightarrow\bket{r}$ on both atoms;
$\bket{0}$ is uncoupled. The interaction shifts $\bket{rr}$ by $B$,
suppressing double excitation in the perfect-blockade limit
$B/\Omax\to\infty$. The collective transition
$\bket{11}\leftrightarrow\bket{W}=(\bket{1r}+\bket{r1})/\sqrt{2}$
has Rabi frequency $\sqrt{2}\Omega(t)$.
(b)~Excitation to the Rydberg manifold and coherent return compete with
decay, shown by the wavy arrow. The orange profile schematically represents
the Rydberg population $\bar p_R(t)$ averaged over the four computational-basis
inputs. Its physical-time integral is the exposure
$\TR=\int_0^T\bar p_R(t)\,\dd t$ for gate duration $T$, and the
dimensionless objective is $C=\Omax\TR$. For irreversible loss from the
computational subspace at single-atom decay rate $\Gamma$, the leading-order
loss probability is $P_{\mathrm{loss}}\approx\Gamma\TR$.
The level spacings and occupation profile are schematic, not quantitative
spectra or a simulated waveform.}}
\label{fig:overview}
\end{figure*}

We express the exact gate dynamics and closure conditions as a polynomial
optimal-control problem. Using semidefinite programming (SDP), we
construct sum-of-squares (SOS) certificates that bound the Rydberg
exposure over all finite gate durations~\cite{Lasserre01,Lasserre08,Parrilo03,Putinar93}.
We rigorously verify the certificates and strengthen the lower bound using constraints imposed by the shared drive. Interval arithmetic then certifies the existence of an exact CZ pulse below the upper endpoint.
{Figure~\ref{fig:overview}(a) shows the shared-drive,
perfect-blockade model, while Fig.~\ref{fig:overview}(b) illustrates
Rydberg occupation, exposure, and decay during the gate.}

Within this model, the certified $0.16\%$ window
leaves little scope for reducing decay loss through further pulse shaping.
Larger reductions require changes in physical parameters, such as
higher laser power or longer-lived atomic states,
or an extension of the control model. Individual atom addressing
and multi-tone dressing warrant investigation as possible extensions.

Sections~\ref{sec:model}--\ref{sec:pulse} present the model and the
certified lower and upper bounds. Section~\ref{sec:discussion} discusses the physical
implications; detailed proofs and computational methods are collected in the
Appendices.

\section{Model and the exposure control problem}
\label{sec:model}

To establish rigorous bounds on the Rydberg exposure, we formulate the gate dynamics as an
optimal control problem. Below, we specify the two-atom Hamiltonian and
terminal conditions for an exact CZ gate up-to local rotations in the perfect-blockade regime,
and prove that operating below peak laser amplitude cannot lower the
exposure cost. This yields an exact reduction to phase-only
dynamics parameterized by active time.

\subsection{Physical model and gate conditions}
\label{sec:model_phys}

Consider two identical neutral atoms, each with computational ground states $\bket0,
\bket1$ and an excited Rydberg state $\bket r$. A single, globally applied laser field
couples $\bket1 \leftrightarrow \bket r$ with a time-dependent, non-negative Rabi
frequency envelope $\Omega(t) \le \Omax$ and laser phase $\chi(t)$
{[Fig.~\ref{fig:overview}(a)]}. In the rotating-wave
approximation and in the frame rotating at the laser carrier frequency, the single-atom
drive Hamiltonian (with $\hbar=1$ and energies expressed in
angular-frequency units) is
\begin{equation}
H_{\mathrm{atom}}(t) \;=\; \frac{\Omega(t)}{2}\ee^{\ii\chi(t)}\bket{r}\bbra{1} \;+\;
\frac{\Omega(t)}{2}\ee^{-\ii\chi(t)}\bket{1}\bbra{r}.
\label{eq:Hatom}
\end{equation}
In the strong-blockade regime ($B/\Omax \to \infty$, where $B$ is the nearest-neighbor
blockade energy), the Rydberg interaction detunes the doubly excited state $\bket{rr}$ far
off resonance, preventing double excitation.
Under a spatially uniform drive, the input $\bket{00}$ is uncoupled
and acquires no dynamical phase. The $\bket{01}$ and $\bket{10}$ sectors have
identical two-level dynamics, $\bket{01}\leftrightarrow\bket{0r}$ and
$\bket{10}\leftrightarrow\bket{r0}$, with Rabi frequency $\Omega(t)$.
The $\bket{11}$ sector couples to the symmetric bright state
$\bket{W}=(\bket{1r}+\bket{r1})/\sqrt{2}$ with Rabi frequency
$\sqrt{2}\Omega(t)$; the antisymmetric dark state remains unpopulated.
We label these two driven cases by $k=1,2$, respectively, counting the atoms
initially in $\bket1$.

Let $\tau = \Omax t$ denote dimensionless laboratory time.
In the $\bket{01}$ sector, let $g_1(\tau),r_1(\tau)$ denote the amplitudes of
$\bket{01},\bket{0r}$, respectively, with identical amplitudes for
$\bket{10},\bket{r0}$. In the $\bket{11}$ sector, let $g_2(\tau),r_2(\tau)$ denote the amplitudes of
$\bket{11},\bket{W}$, respectively.
The normalization is $|g_k(\tau)|^2+|r_k(\tau)|^2=1$ for $k\in\{1,2\}$.
The normalized physical drive is
\begin{equation}
\begin{gathered}
v(\tau) \;:=\; \frac{\Omega(t)\ee^{-\ii\chi(t)}}{\Omax} \;=\; \rho(\tau)\ee^{-\ii\chi(t)},
\end{gathered}
\label{eq:physical_control}
\end{equation}
where $\rho(\tau) = \Omega(t)/\Omax \leq 1$ is the normalized amplitude. In normalized laboratory time, the
Schrödinger equation for each sector is
\begin{equation}
\ii \frac{\dd}{\dd \tau} \begin{pmatrix} g_k \\ r_k \end{pmatrix}
\;=\;
\frac{\sqrt{k}}{2}
\begin{pmatrix}
0 & v(\tau) \\
v^*(\tau) & 0
\end{pmatrix}
\begin{pmatrix} g_k \\ r_k \end{pmatrix},
\qquad k \in \{1, 2\}.
\label{eq:schrod_lab}
\end{equation}
Initially, the atoms reside entirely in their computational ground states:
\begin{equation}
g_1(0)=g_2(0)=1,\qquad r_1(0)=r_2(0)=0.
\label{eq:initial_cond}
\end{equation}
At population return, $r_1(\tau_f)=r_2(\tau_f)=0$, normalization gives
$g_k(\tau_f)=\ee^{\ii\Phi_k}$, and the computational gate is
\begin{equation*}
\operatorname{diag}\bigl(1,\ee^{\ii\Phi_1},\ee^{\ii\Phi_1},\ee^{\ii\Phi_2}\bigr).
\end{equation*}
After identical single-qubit $Z$ rotations, this gate is equivalent, up to an
overall phase, to $\operatorname{diag}\bigl(1,1,1,\ee^{\ii(\Phi_2-2\Phi_1)}\bigr)$.
It therefore realizes CZ precisely when
$\Phi_2-2\Phi_1\equiv\pi\pmod{2\pi}$~\cite{Levine19}.
Together, these requirements give the closure conditions:
\begin{equation}
r_1(\tau_f) \;=\; r_2(\tau_f) \;=\; 0,
\qquad
g_2(\tau_f) \;=\; - g_1(\tau_f)^2.
\label{eq:CZ_cond}
\end{equation}

For the computational-basis inputs, the instantaneous Rydberg population is $0$
for $\bket{00}$, $|r_1|^2$ for each of $\bket{01}$ and $\bket{10}$, and $|r_2|^2$ for
$\bket{11}$. We collect the amplitudes of the two driven sectors into the
vector $\xv=(g_1,r_1,g_2,r_2)\in\mathbb{C}^4$. The two-sector state space consists of these
vectors satisfying $|g_k|^2+|r_k|^2=1$ for $k=1,2$. Here $\xv$ without an argument
denotes a point of this state space and $\xv(\tau)$ the point reached along a pulse.
The input-averaged exposure cost is $c(\xv):=(2|r_1|^2+|r_2|^2)/4$, and the total
dimensionless exposure {[Fig.~\ref{fig:overview}(b)]} is
\begin{equation}
C_{\mathrm{lab}}[v] \;=\; \Omax\TR \;=\; \int_0^{\tau_f} c(\xv(\tau))\,\dd\tau.
\label{eq:objective_lab}
\end{equation}
Let $C_*$ denote the infimum of $C_{\mathrm{lab}}[v]$ over all
admissible exact CZ pulses and finite gate durations. Admissible pulses have
measurable controls $|v|\le1$ and satisfy the dynamics~\eqref{eq:schrod_lab},
initial condition~\eqref{eq:initial_cond}, and CZ target~\eqref{eq:CZ_cond}.
We allow arbitrary phase modulation, so our lower bound also applies
to bandwidth-limited pulses governed by the same dynamics.

\subsection{Shared phase control in active time}
\label{sec:reduction}

The full variational problem entails co-optimizing the amplitude envelope $\rho(\tau) \in
[0, 1]$ and the phase profile $\chi(\tau)$. However, throttling the laser amplitude below
maximum never reduces the exposure:

\begin{lemma}[Maximum-amplitude reduction and active time]
\label{lem:bang}
Every admissible control with $|v|\le1$ can be replaced by a
control with $|v|=1$ that produces the same gate in the same or a shorter time,
with no greater exposure. Thus restricting to $|v|=1$ leaves $C_*$ unchanged.
\end{lemma}
\begin{proof}
For any admissible physical control, define the accumulated active time
\begin{equation}
s(\tau) \;:=\; \int_0^\tau \rho(q)\,\dd q
\;=\; \int_0^t \Omega(t')\,\dd t',
\qquad S:=s(\tau_f).
\label{eq:active_time}
\end{equation}
Because the state velocity in~\eqref{eq:schrod_lab} is proportional to $\rho(\tau)$,
reparameterizing non-idling intervals ($\rho > 0$) by active time and setting $u(s) :=
\ee^{-\ii\chi(\tau(s))}$, where $\tau(s)$ is the inverse time map after removing
idling, yields the full-amplitude dynamics
\begin{equation}
\ii \frac{\dd}{\dd s}\begin{pmatrix}g_k\\r_k\end{pmatrix}
\;=\; \frac{\sqrt{k}}{2}
\begin{pmatrix}0&u(s)\\u^*(s)&0\end{pmatrix}
\begin{pmatrix}g_k\\r_k\end{pmatrix},
\qquad k \in \{1, 2\}.
\label{eq:schrod}
\end{equation}
The unit-modulus control $u(s)$ ($|u|=1$) thus reproduces the same populations and phases
at maximum available Rabi rate ($\rho = 1$). Any interval with $\rho = 0$
corresponds to stationary idling: if the system resides entirely in the computational
ground state ($c(\xv)=0$), idling produces zero gate evolution at zero cost; if any
Rydberg population is present ($c(\xv)>0$), idling incurs spontaneous decay without
advancing the gate dynamics. Because $c(\xv) \ge 0$ and $\dd s = \rho(\tau)\,\dd\tau \le
\dd\tau$, the exposure of the active-time waveform satisfies
\begin{equation}
\begin{split}
C[u] &:= \int_0^S c(\xv(s))\,\dd s \\
&= \int_0^{\tau_f} c(\xv(\tau))\rho(\tau)\,\dd\tau \;\le\; C_{\mathrm{lab}}[v].
\end{split}
\label{eq:compression_cost}
\end{equation}
Because maximum-amplitude waveforms form a subset of all admissible physical controls,
their infimum cannot be strictly less than $C_*$. Conversely, the
mapping~\eqref{eq:compression_cost} associates each admissible control with a
full-amplitude waveform of equal or lower exposure, showing that their infimum cannot
exceed $C_*$. The two infima must therefore coincide.
\end{proof}

Henceforth, we work exclusively in active time $s \in [0, S]$ governed
by~\eqref{eq:schrod} with unit-modulus control $u(s) = \ee^{-\ii\chi(s)}$, subject to the
initial state~\eqref{eq:initial_cond} and terminal conditions~\eqref{eq:CZ_cond}. The
active-time exposure functional becomes
\begin{equation}
\begin{gathered}
C[u] \;=\; \frac{2E_1+E_2}{4}, \\
E_k \;:=\; \int_0^S |r_k(s)|^2\,\dd s.
\end{gathered}
\label{eq:objective}
\end{equation}
The physical infimum $C_*$ is therefore identically the infimum of $C[u]$ over all
measurable unit-modulus controls $u(s)$ and all finite durations $S$.

The shared drive ties both sectors to the same phase profile $u(s)$, with
relative coupling strengths $1$ and $\sqrt{2}$. If the sectors instead had independent
controls, their exposure infimum would vanish (Appendix~\ref{sec:nogo}).

To see why the shared drive enforces a positive exposure cost, compare the
blockaded collective state with the state of two noninteracting atoms under
the same drive. Both start in $\bket{11}$. At exact CZ closure, their final
states differ by a minus sign because $g_2=-g_1^2$, so their overlap amplitude must
change from $+1$ to $-1$. The common part of their Hamiltonians cancels
from the derivative of this overlap. The magnitude of the remaining contribution
is bounded in terms of the Rydberg populations that also determine the exposure rate.
Appendix~\ref{sec:analytical} bounds the overlap change per unit accumulated
exposure. The required change from $+1$ to $-1$ therefore has a positive
exposure cost, regardless of how long the gate takes.

The analytical bound retains only part of the coupled dynamics and is
too weak to determine whether pulses with $C\approx2.936$ are close to
optimal. Section~\ref{sec:moments_sos} uses a moment-SOS
hierarchy with the full dynamics to tighten the certified bound.

\section{Lower bound from the cumulative-cost dual}
\label{sec:moments_sos}

We begin by writing the active-time infimum of Sec.~\ref{sec:reduction} as a
standard program in the pulse, the duration, and the trajectory. A real
polynomial on the two-sector state space then supplies a lower bound that does
not depend on any particular pulse. SOS certificates convert the
resulting positivity constraints into a hierarchy of SDPs;
symmetry reduces their size; and a cumulative-exposure partition narrows the
remaining gap. An exact dual certificate over $\mathbb{Q}(\sqrt{2})$ together with a
physical-slack extension yields the lower bound in Eq.~\eqref{eq:headline}.

\subsection{Sum-of-squares formulation of the lower bound}
\label{sec:occupation}
\label{sec:unsplit_progression}

We write the active-time minimization of $C[u]$ over unit-modulus
pulses and finite durations (Sec.~\ref{sec:reduction}) as an optimal control problem
with polynomial dynamics and constraints. Let $\xv=(g_1,r_1,g_2,r_2)$ collect the
amplitudes of the two driven sectors under the common control $u$, as in
Sec.~\ref{sec:model_phys}.
Multiplying~\eqref{eq:schrod} by $-\ii$ and stacking the two sectors then
gives $\dd \xv/\dd s=f(\xv,u)$, with
\begin{equation}
f(\xv,u)=-\frac{\ii}{2}\bigl(u\,r_1,\ \bar u\,g_1,\ \sqrt2\,u\,r_2,\ \sqrt2\,\bar
u\,g_2\bigr).
\label{eq:vector_field}
\end{equation}
Each component of $f$ is a polynomial of total degree two in
$(\xv,u)$ and their complex conjugates. The running cost
$c(\xv)=(2|r_1|^2+|r_2|^2)/4$ of~\eqref{eq:objective_lab} is quadratic.
For any admissible pulse, the state--control pair
$(\xv(s),u(s))$ lies in the set
\begin{equation}
\begin{gathered}
\mathcal K:=\mathcal{X}\times\{u:|u|^2=1\},\\
\mathcal{X}:=\{\xv\in\mathbb{C}^4:|g_1|^2+|r_1|^2=1,\ |g_2|^2+|r_2|^2=1\},
\end{gathered}
\label{eq:state_set}
\end{equation}
and the state trajectory
$\xv(s)$ starts at $\xv_0=(1,0,1,0)$ and ends on the CZ terminal orbit
\begin{equation}
\mathcal{X}_{\mathrm{CZ}}:=\left\{\xv\in\mathcal{X}:r_1=r_2=0,\ g_2=-g_1^2\right\},
\label{eq:cz_terminal_set}
\end{equation}
where every point on this orbit corresponds to a CZ gate up to identical single-qubit $R_z$ rotations.

The resulting optimization program seeks to minimize the exposure $C[u]$ over the unit-modulus pulse $u(s)$, finite duration $S$, and trajectory $\xv(s)$:
\begin{equation}
\begin{aligned}
C_* \;:=\; \inf_{u(\cdot),\,S,\,\xv(\cdot)}
& \quad C[u]=\int_0^S c\bigl(\xv(s)\bigr)\,\dd s \\
\text{subject to:}
& \quad \dd \xv/\dd s = f\bigl(\xv(s),u(s)\bigr), \quad 0\le s\le S, \\
& \quad \xv(0)=\xv_0=(1,0,1,0), \\
& \quad \bigl(\xv(s),u(s)\bigr)\in\mathcal K \quad \text{for all } s, \\
& \quad \xv(S)\in\mathcal{X}_{\mathrm{CZ}}.
\end{aligned}
\label{eq:exposure}
\end{equation}
The sets $\mathcal K$ and $\mathcal{X}_{\mathrm{CZ}}$ are defined by
the polynomial equalities in Eqs.~\eqref{eq:state_set}
and~\eqref{eq:cz_terminal_set}, respectively, and $\mathcal K$ is bounded.

Direct pulse optimization can identify low-exposure candidates,
but a candidate alone does not establish a lower bound valid for every admissible pulse.
To bound $C_*$ in Eq.~\eqref{eq:exposure} from below, we seek a
continuously differentiable function $W(\xv)$ that gives a lower bound on the
exposure required to reach the CZ terminal orbit from $\xv$. On the terminal
orbit, no further exposure is required, so we impose
\begin{equation}
W(\xv)\le 0 \qquad\text{for all }\xv\in\mathcal{X}_{\mathrm{CZ}}.
\label{eq:W_term}
\end{equation}
With the terminal condition~\eqref{eq:W_term}, it suffices to
require that the accumulated exposure plus $W(\xv(s))$ be nondecreasing along
every admissible trajectory. This gives the differential condition
\begin{equation}
\frac{\dd W}{\dd s}\ge -c.
\label{eq:W_drift}
\end{equation}
Integrating Eq.~\eqref{eq:W_drift} over $s\in[0,S]$ and applying
the terminal condition~\eqref{eq:W_term} gives
\begin{equation*}
C[u]\ge W(\xv_0)-W(\xv(S))\ge W(\xv_0).
\end{equation*}
To express these conditions as polynomial nonnegativity constraints,
we choose $W=-n-V$, where $n$ is a real constant and $V$ is a real polynomial in
the real and imaginary parts of $\xv$. Because $\dd W/\dd s=-\nabla V\cdot f$,
condition~\eqref{eq:W_term} is $p_{\mathrm{term}}\ge0$ and
condition~\eqref{eq:W_drift} is $p_{\mathrm{drift}}\ge0$, where
\begin{equation}
\begin{gathered}
p_{\mathrm{drift}}(\xv,u):=c(\xv)-\nabla V(\xv)\cdot f(\xv,u),\\
p_{\mathrm{term}}(\xv):=V(\xv)+n,
\end{gathered}
\label{eq:barrier_polys}
\end{equation}
where $\nabla V\cdot f$ is the directional derivative of $V$ along $f(\xv,u)$.
Because $V$ and $f$ are polynomials, so is $p_{\mathrm{drift}}$. We call
$p_{\mathrm{drift}}\ge0$ on $\mathcal K$ the drift inequality and
$p_{\mathrm{term}}\ge0$ on $\mathcal{X}_{\mathrm{CZ}}$ the terminal inequality.

Each pair $(V,n)$ satisfying the drift and terminal inequalities
gives $C[u]\ge W(\xv_0)=-n-V(\xv_0)$ for every admissible pulse.
At relaxation order $d$, we restrict $V$ to degree at most $2d-1$,
so that the drift polynomial has degree at most $2d$.
We then take the supremum over $V$ and $n$:
\begin{equation}
\begin{aligned}
\sup_{V,\,n}
& \quad W(\xv_0)=-n-V(\xv_0) \\
\text{subject to:}
& \quad \deg V\le 2d-1,\quad n\in\mathbb{R}, \\
& \quad p_{\mathrm{drift}}(\xv,u)\ge0\quad\text{for all }(\xv,u)\in\mathcal K, \\
& \quad p_{\mathrm{term}}(\xv)\ge0\quad\text{for all }\xv\in\mathcal{X}_{\mathrm{CZ}}.
\end{aligned}
\label{eq:pop}
\end{equation}
This is an optimization problem in the coefficients of $V$ and the scalar $n$.
The state and control symbols $g_k$, $r_k$, and $u$, together
with their complex conjugates, are arguments of the polynomials. By the
preceding integration argument, each feasible pair $(V,n)$ bounds every
admissible finite-duration pulse. The two inequalities
must hold throughout their respective domains, so Eq.~\eqref{eq:pop} contains
infinitely many pointwise constraints.

We impose a sufficient condition for these inequalities: each polynomial
admits a sum-of-squares representation on its domain. At fixed degree, this requirement
can be expressed as a finite semidefinite program.
We first reduce monomials using their defining equations.
On $\mathcal K$,
\begin{equation*}
\bar u\,u=1,\qquad \bar g_k g_k=1-\bar r_k r_k\quad (k=1,2).
\end{equation*}
After these substitutions a monomial never carries a
simultaneous barred and unbarred power of $u$, $g_1$, or $g_2$.
Let $z$ be the column vector of all reduced monomials of total degree at most
$d$ in $u,g_1,r_1,g_2,r_2$ and their complex conjugates. On $\mathcal{X}_{\mathrm{CZ}}$, we impose the two state normalizations together with
\begin{equation*}
r_1=r_2=0,\qquad g_2=-g_1^2.
\end{equation*}
The terminal relations reduce the state variables to $g_1$ and $\bar g_1$,
with $g_1\bar g_1=1$. We may therefore choose
\begin{equation*}
w=\bigl(1,g_1,\bar g_1,\ldots,g_1^d,\bar g_1^{\,d}\bigr)^{\mathsf T}.
\end{equation*}
We construct the real-valued polynomial $V(\xv)$ from the reduced monomials in
the state amplitudes and their complex conjugates; it does not depend on the
control $u$. A representation $z^*Qz$ (or $w^*Qw$) with a Hermitian
positive semidefinite (PSD) Gram matrix $Q\succeq0$ is a sum of squared moduli
of polynomials and is therefore nonnegative.
After the symmetry reduction in Sec.~\ref{sec:sos_reductions}, the matrix $Q$ can be taken real symmetric.
Matching coefficients after these reductions gives a finite set of linear
equations in the coefficients of $V$, the scalar $n$, and the entries of the
two Gram matrices. The terminal substitutions apply only to the terminal
identity; the objective evaluates $V$ at $\xv_0$ without those substitutions. We obtain the SDP~\cite{Lasserre01,Parrilo03,Putinar93}:
\begin{equation}
\begin{aligned}
\sup_{V,\,n,\,Q}
& \quad -n-V(\xv_0) \\
\text{subject to:}
& \quad p_{\mathrm{drift}}
\equiv z^*Q_{\mathrm{drift}}z
\quad (\mathcal K), \\
& \quad Q_{\mathrm{drift}}\succeq0, \\
& \quad p_{\mathrm{term}}
\equiv w^*Q_{\mathrm{term}}w
\quad (\mathcal{X}_{\mathrm{CZ}}), \\
& \quad Q_{\mathrm{term}}\succeq0.
\end{aligned}
\label{eq:sdp}
\end{equation}
Any feasible point of~\eqref{eq:sdp} is feasible for~\eqref{eq:pop}. We call this formulation \emph{unsplit}: it uses a single polynomial
$V(\xv)$ that depends only on the state, without dividing the cumulative-exposure range into intervals.

The certificate conditions in Eq.~\eqref{eq:sdp} provide the starting point for the
exposure-partition construction in Sec.~\ref{sec:exposure_partition}.
Section~\ref{sec:sos_reductions} explains how symmetry reduces the computational
cost of finding the Gram matrices; its basis transformations can be skipped
on a first reading of the lower-bound argument. Appendix~\ref{app:split_certificate}
specifies the bases and exact checks needed to reproduce the certificates.

\subsection{Symmetry reduction of the unsplit program}
\label{sec:sos_reductions}

The sizes of Gram matrices in Eq.~\eqref{eq:sdp} grow rapidly with $d$.
To reduce them, we use two phase rotations and a reflection that preserve
the equations of motion, the running cost $c$, and the sets $\mathcal K$ and
$\mathcal{X}_{\mathrm{CZ}}$.
Their preservation of the equations of motion and the CZ terminal conditions
can be checked directly in Eqs.~\eqref{eq:schrod} and~\eqref{eq:CZ_cond}, respectively.
\begin{enumerate}
\setlength{\itemsep}{3pt}
\setlength{\parsep}{0pt}
\item[(i)] \begin{minipage}[t]{\dimexpr\columnwidth-\leftmargin\relax}
\textbf{{Optical phase symmetry.}}
A constant phase shift of the drive is paired with an opposite phase rotation
of the Rydberg amplitudes:
\[
u\mapsto\ee^{\ii\theta}u,\qquad
g_k\mapsto g_k,\qquad
r_k\mapsto\ee^{-\ii\theta}r_k.
\]
\end{minipage}
\item[(ii)] \begin{minipage}[t]{\dimexpr\columnwidth-\leftmargin\relax}
\textbf{{Qubit phase symmetry.}}
A common local $Z$ rephasing, extended to give $\bket{r}$ the same phase as
$\bket{1}$, acts as
\[
\begin{aligned}
u&\mapsto u,\\
(g_1,r_1)&\mapsto\ee^{\ii\alpha}(g_1,r_1),\\
(g_2,r_2)&\mapsto\ee^{2\ii\alpha}(g_2,r_2).
\end{aligned}
\]
The phase weights $1$ and $2$ preserve the CZ relation $g_2=-g_1^2$.
\end{minipage}
\item[(iii)] \begin{minipage}[t]{\dimexpr\columnwidth-\leftmargin\relax}
\textbf{Reflection.}
Complex conjugation of the state amplitudes is accompanied by the drive
transformation
\[
u\mapsto-\bar u,\qquad
g_k\mapsto\bar g_k,\qquad
r_k\mapsto\bar r_k.
\]
This transformation preserves forward-time evolution and sends the laser phase
to $\chi\mapsto\pi-\chi$ modulo $2\pi$.
\end{minipage}
\end{enumerate}
These operations generate the symmetry group $G$.
The two phase rotations group basis monomials by their phase response;
the reflection then allows the resulting Gram blocks to be chosen real
symmetric.

We restrict $V$ to polynomials invariant under the two phase rotations,
retaining the constraints of Eq.~\eqref{eq:pop}.
The dynamics and cost respect these symmetries, so the drift polynomial is also invariant.

The \emph{charge} of a monomial $m$ is the integer pair
$q(m)=(q_\theta,q_\alpha)$ that determines its phase under rotations (i) and (ii):
\[
m\mapsto\ee^{\ii(q_\theta\theta+q_\alpha\alpha)}m.
\]
Multiplying monomials adds their charges, while complex conjugation
reverses the sign. The \emph{total charge} of a product is therefore the
sum of the charges of its factors, including these signs for conjugated
factors. A monomial is invariant under both rotations exactly when this
pair is $(0,0)$. For example, $g_1$ and $ur_1$ both have charge $(0,1)$:
they acquire the same phase even though their degrees differ. Such
\emph{equal-charge} monomials have an invariant conjugate product,
here $\bar g_1ur_1$.

In the Gram representation $z^*Q_{\mathrm{drift}}z$, the entry
$(Q_{\mathrm{drift}})_{ij}$ multiplies $\bar z_i z_j$, whose charge is
$q(z_j)-q(z_i)$. Averaging over both rotations removes entries joining
different charges while preserving the invariant drift polynomial and
positive semidefiniteness~\cite{Gatermann04,Riener13}. Ordering the basis
by charge then makes $Q_{\mathrm{drift}}$ block diagonal. Each
\emph{charge block} contains the rows and columns for one common charge;
its dimension is the number of basis monomials with that charge.
Thus one large PSD constraint becomes several smaller PSD constraints.
Appendix~\ref{app:terminal_quotient} gives the explicit averaging formula
and the example shown in Fig.~\ref{fig:method}.

The reflection in (iii) maps a feasible polynomial $V(\xv)$ to another
feasible polynomial $V(\bar{\xv})$. Their average is reflection
invariant and remains feasible. Since $\bar{\xv}_0=\xv_0$, this averaging
leaves the objective unchanged.
Multiplying each SOS basis monomial with odd total degree in $u$ and $\bar{u}$ by $\ii$
allows the charge blocks to be chosen real symmetric.

On the zero-cost {set} $F=\{r_1=r_2=0\}$, the drift inequality for both $u$
and $-u$ forces $p_{\mathrm{drift}}=0$, so every SOS factor vanishes there
(Appendix~\ref{subsec:positivity_cones}). Compressing each charge block to a basis of
these factors further reduces the Gram matrices. The resulting block dimensions
are listed in Appendix Table~\ref{tab:sos_progression}.

Increasing the relaxation order to $11$ yields an unsplit numerical objective
of approximately $2.9047$, still about $0.0311$ below the certified upper bound.
The number of PSD entries grows by more than a factor of $250$ between orders $6$ and $11$
(Appendix Table~\ref{tab:sos_progression}). Going to higher orders is therefore
computationally expensive. Only the order-$8$ unsplit solution was converted
into an exact certificate; floating-point objectives alone are not certified bounds.
Within the invariant polynomial class,
symmetry and face reduction reduce the matrix size without weakening
the SOS representation. The next subsection introduces a cumulative-exposure coordinate
and partitions its range, yielding a tighter bound at order~$6$.

\subsection{Cumulative-exposure partition}
\label{sec:exposure_partition}

Improving the bound by increasing the relaxation order is computationally expensive.
The optimal remaining exposure depends only on $\xv$; the difficulty is
obtaining a tight lower bound using a low-degree polynomial.
At a fixed polynomial degree, we introduce a second argument,
the budget coordinate $X$: the fraction of a known exposure ceiling $U$
already spent. Partitioning the range of $X$ and using a low-degree polynomial
in $(X,\xv)$ on each cell can tighten the lower bound.

That ceiling is not a new constraint on the laser.
Sec.~\ref{sec:interval_closure} uses interval arithmetic to certify an exact CZ
gate with exposure below $C_{\mathrm{up}}$. We choose a rational ceiling
$U>C_{\mathrm{up}}$, whose value is specified in Appendix~\ref{app:order7_certificate}.
Pulses with $C[u]\ge U$ cannot improve
on this gate and can be excluded from the minimization. For every remaining
pulse, we define the normalized cumulative exposure
\begin{equation}
X(s):=\frac1U\int_0^s c\bigl(\xv(s')\bigr)\,\dd s'\in[0,1)
\label{eq:exposure_odometer}
\end{equation}
The coordinate $X$ increases whenever Rydberg states are
populated, even if the state does not approach the CZ target. Using $X$ gives
the compact domain $[0,1]\times\mathcal K$ for the SOS
representation. The active time $s$ has no uniform upper bound
over admissible pulses.

However, making the domain compact does not by itself ensure
that a single low-degree polynomial can give a tight lower bound.
For the pulses retained above, $X=0$ initially and $X\approx1$ at
gate closure. Near both endpoints, $r_k\approx0$, so the small running cost
limits the decrease of the estimate through $\dd W/\dd s\ge-c$.
A tight estimate should be large near $\xv_0$ and close to zero near the CZ
target. We therefore partition $[0,1]$ at rational knots and assign a separate
polynomial to each cell. Because $X$ is nondecreasing, trajectories can cross
cell boundaries only from left to right. Cell $j$ has width $w_j$ and a local
coordinate $\xi_j\in[0,1]$,
\begin{equation}
\begin{gathered}
0=a_0<a_1<\cdots<a_J=1,\\
w_j=a_j-a_{j-1},\qquad\xi_j=\frac{X-a_{j-1}}{w_j}.
\end{gathered}
\label{eq:split_grid}
\end{equation}
Within cell $j$, we write $\xi=\xi_j$, with $\xi=0$ at the left
boundary and $\xi=1$ at the right boundary. The remaining-exposure estimate is a polynomial
$W_j(\xi,\xv)=-n-V_j(\xi,\xv)$, with the same shift $n$ as in
Sec.~\ref{sec:occupation}.

\begin{table}[t]
\centering
\caption{Numerical relaxation values and certified lower bounds
for the unsplit and cumulative-exposure-split moment-SOS formulations.
Starred ($\ast$) entries are certified exact algebraic dual bounds over
$\mathbb{Q}(\sqrt{2})$; the daggered ($\dagger$) entry denotes the exact
physical-slack extension built upon the order-$7$ dual certificate.
Unmarked entries are floating-point SDP solver objectives
$\widehat{L}_d$ and are not certified lower bounds.
Numerical entries are rounded for display; the final certified lower bound
is stated exactly in Theorem~\ref{thm:slack_strengthened}.
Unsplit solver dimensions appear in Appendix Table~\ref{tab:sos_progression}.
}
\label{tab:sos_comparison}
\vspace{0.3em}
\normalsize
\begin{tabular*}{\columnwidth}{@{\extracolsep{\fill}}cccc@{}}
\toprule
Order $d$ & Cells & Bound / obj. & Gap to $C_{\mathrm{up}}$ \\
\midrule
\multicolumn{4}{@{}l@{}}{\emph{Unsplit}} \\
6  & --- & $2.8625$ & $0.0734$ \\
8  & --- & $2.8825^\ast$ & $0.0534$ \\
11 & --- & $2.9047$ & $0.0311$ \\
\midrule
\multicolumn{4}{@{}l@{}}{\emph{Split}} \\
6 & 5 & $2.9300^\ast$ & $0.0059$ \\
6 & 6 & $2.9300^\ast$ & $0.0059$ \\
7 & 5 & $2.9305^\ast$ & $0.0054$ \\
\midrule
\multicolumn{4}{@{}l@{}}{\emph{Slack extension}} \\
7 (base) & 5 & $\approx\mathbf{2.9313}^\dagger$ & $\mathbf{0.0046}$ \\
\bottomrule
\end{tabular*}
\end{table}

As in Sec.~\ref{sec:occupation}, we require $\dd W_j/\dd s\ge-c$.
Here $V_j$ depends on both $\xv$ and $\xi$, with
$\dd\xi_j/\dd s=c(\xv)/(Uw_j)$. The chain rule therefore adds a
$\xi$-derivative term, giving the drift polynomial
\begin{equation}
\begin{gathered}
p_{\mathrm{drift},j}
:=c-\nabla_{\xv} V_j\cdot f-\frac{c}{Uw_j}\,\partial_\xi V_j,\\
L_j:=f\cdot\nabla_{\xv}+\frac{c}{Uw_j}\,\partial_\xi,
\end{gathered}
\label{eq:split_generator_mass}
\end{equation}
with $p_{\mathrm{drift},j}=c-L_jV_j$. The drift inequality is
$p_{\mathrm{drift},j}\ge0$ on $[0,1]\times\mathcal K$.
Crossing a cell boundary introduces no jump in the accumulated
exposure. We therefore require $W_j(1,\xv)\le W_{j+1}(0,\xv)$, or equivalently
$V_j(1,\xv)\ge V_{j+1}(0,\xv)$ on $\mathcal X$, so that the estimate does not
jump downward. In every cell, we impose $W_j\le0$ on the CZ set.
Maximizing the initial lower bound over polynomials $V_j$ of total degree at most
$2d-1$ gives
\begin{equation}
\begin{aligned}
\sup_{V_1,\dots,V_J,\,n}
& \quad W_1(0,\xv_0)=-n-V_1(0,\xv_0) \\
\text{subject to:}
& \quad \deg V_j\le 2d-1,\quad n\in\mathbb{R}, \\
& \quad p_{\mathrm{drift},j}(\xi,\xv,u)\ge0
\quad\text{on }[0,1]\times\mathcal K,\\
& \quad\qquad j=1,\dots,J, \\
& \quad V_j(1,\xv)-V_{j+1}(0,\xv)\ge0
\quad\text{on }\mathcal{X},\\
& \quad\qquad j=1,\dots,J-1, \\
& \quad V_j(\xi,\xv)+n\ge0
\quad\text{on }[0,1]\times\mathcal{X}_{\mathrm{CZ}},\\
& \quad\qquad j=1,\dots,J.
\end{aligned}
\label{eq:split_pop}
\end{equation}

To obtain a finite SDP, we impose the following sufficient SOS conditions,
extending Eq.~\eqref{eq:sdp}:
\begin{equation*}
\begin{gathered}
p_{\mathrm{drift},j}
\equiv\sigma^{\mathrm{drift}}_{j,0}
+\xi(1-\xi)\sigma^{\mathrm{drift}}_{j,1}
\quad(\mathcal K),\\[2pt]
V_j(1,\xv)-V_{j+1}(0,\xv)
\equiv\sigma^{\mathrm{int}}_j(\xv)
\quad(\mathcal X),\\[2pt]
\begin{aligned}
V_j(\xi,\xv)+n
&\equiv\sigma^{\mathrm{term}}_{j,0}\\
&\quad+\xi(1-\xi)\sigma^{\mathrm{term}}_{j,1}
\quad(\mathcal X_{\mathrm{CZ}}).
\end{aligned}
\end{gathered}
\end{equation*}
Here each $\sigma$ denotes an SOS polynomial in the corresponding reduced variables.
Their degree bounds, bases, and cellwise charge blocks are specified in
Appendix~\ref{subsec:positivity_cones}. The proof of Theorem~\ref{thm:split_exact}
shows that each feasible solution of Eq.~\eqref{eq:split_pop} gives
\begin{equation*}
C_*\ge-n-V_1(0,\xv_0).
\end{equation*}

For the order-$7$ certificate, we use the five-cell grid
\begin{equation}
(a_0,\ldots,a_5)=\left(0,\frac1{36},\frac13,\frac23,\frac{35}{36},1\right).
\label{eq:primary_split_grid}
\end{equation}
The two narrow end cells are each rescaled to $\xi\in[0,1]$
and assigned a separate
polynomial estimate; the middle cells cover wider exposure intervals.
The next subsection constructs an exact certificate from a
floating-point solution of the SOS program.

\subsection{Exact certificate over \texorpdfstring{$\mathbb{Q}(\sqrt{2})$}{Q(sqrt(2))}}
\label{sec:split_exact}

Even a highly accurate floating-point solution does not by itself
provide a rigorous SOS certificate, because the constraints may not hold exactly.
We therefore reconstruct and verify an exact certificate.
To choose the coefficient field for exact reconstruction, we examine
the drift polynomial in Eq.~\eqref{eq:split_generator_mass}, which contains the vector
field in Eq.~\eqref{eq:vector_field}:
\begin{equation}
\begin{gathered}
L_jV_j
= f\cdot\nabla_{\xv} V_j+\frac{c(\xv)}{Uw_j}\,\partial_\xi V_j,\\
f=-\frac{\ii}{2}\bigl(u\,r_1,\ \bar u\,g_1,\ \sqrt2\,u\,r_2,\ \sqrt2\,\bar u\,g_2\bigr).
\end{gathered}
\label{eq:Lj_sqrt2}
\end{equation}
The factor $\sqrt{2}$ in $f$ comes from the two-atom coupling.
The running cost $c$ has rational coefficients, and both $U$ and the cell
widths $w_j$ are rational. After the basis rephasing described in
Sec.~\ref{sec:sos_reductions}, we express the certificate coefficients in the
quadratic field $\mathbb{Q}(\sqrt{2})=\{a+b\sqrt{2}:a,b\in\mathbb{Q}\}$:
\begin{equation}
V_j(\xi,\xv)
=\sum_{\alpha}\bigl(a_{j\alpha}+b_{j\alpha}\sqrt{2}\bigr)\,m_{\alpha}(\xi,\xv),
\qquad a_{j\alpha},b_{j\alpha}\in\mathbb{Q},
\label{eq:Vj_qsqrt2}
\end{equation}
where $\{m_{\alpha}(\xi,\xv)\}_{\alpha}$ is the chosen reduced monomial basis in the cell
coordinate $\xi$ and state amplitudes $\xv$, indexed by $\alpha$.
Numerical SDPs supply floating-point seeds for those
coefficients, the scalar $n$, and the Gram matrices.
We reconstruct a certificate over $\mathbb{Q}(\sqrt{2})$, and then use
exact arithmetic to verify its polynomial identities and the positive
semidefiniteness of its Gram matrices~\cite{Rump06} (Appendix~\ref{app:auditor}).

\begin{theorem}[Exact five-cell split dual]
\label{thm:split_exact}
For globally driven CZ gates in the perfect-blockade model
defined in Sec.~\ref{sec:model}, with $\Omega(t)\le\Omax$, the exposure infimum satisfies
\begin{equation}
C_* \;\ge\; L:=2.9305.
\label{eq:split_bound}
\end{equation}
\end{theorem}
\begin{proof}
We integrate the drift inequality cell by cell and apply the interface
and terminal conditions to establish a lower bound on the total exposure.
Fix any competitive trajectory with $C<U$ and the five-cell
partition~\eqref{eq:primary_split_grid}.
Let $j_*$ be the index of the cell containing $X(S)\in[0,1)$,
choosing the left cell if the endpoint lies on an interface. Write
$0=s_0<\cdots<s_{j_*-1}<s_{j_*}=S$, where $s_j$ for $1\le j<j_*$ is the active
time at which the trajectory crosses $X=a_j$. The exact
certificate supplies a feasible point of~\eqref{eq:split_pop}: real polynomials
$V_1,\ldots,V_5$ of degree at most $13$ and a scalar $n\in\mathbb{Q}(\sqrt{2})$
such that $c-L_jV_j\ge0$ throughout each cell, $V_j(1,\cdot)\ge V_{j+1}(0,\cdot)$
at every interface, and $V_j+n\ge0$ on the terminal orbit
(Appendix~\ref{app:order7_certificate}). Integrating $\dd V_j/\dd s=L_jV_j\le c(\xv)$
along each traversed segment and summing yields
\begin{equation}
\begin{aligned}
C
&=\sum_{j=1}^{j_*-1}\int_{s_{j-1}}^{s_j}c\bigl(\xv(s)\bigr)\,\dd s\\
&\qquad+\int_{s_{j_*-1}}^{S}c\bigl(\xv(s)\bigr)\,\dd s\\
&\ge\sum_{j=1}^{j_*-1}\Bigl[V_j\bigl(1,\xv(s_j)\bigr)-V_j\bigl(0,\xv(s_{j-1})\bigr)\Bigr]\\
&\qquad+\Bigl[V_{j_*}\bigl(\xi(S),\xv(S)\bigr)-V_{j_*}\bigl(0,\xv(s_{j_*-1})\bigr)\Bigr]\\
&\ge V_{j_*}\bigl(\xi(S),\xv(S)\bigr)-V_1(0,\xv_0)\\
&\ge -n-V_1(0,\xv_0)\\
&=L.
\end{aligned}
\label{eq:split_weak_duality}
\end{equation}
Dropping the nonnegative interface terms
$V_j(1,\xv)-V_{j+1}(0,\xv)$ at each crossed interface yields the second inequality.
The terminal condition gives $V_{j_*}(\xi(S),\xv(S))\ge -n$,
yielding the third inequality. The final equality is the certified objective value.
The three nonnegativity conditions are certified by SOS representations
over $\mathbb{Q}(\sqrt{2})$ (Appendix~\ref{subsec:positivity_cones}). The exact audit
verifies the polynomial identities and the positive semidefiniteness of the
Gram matrices without floating-point tolerances (Appendix~\ref{app:order7_certificate}).
Trajectories with $C\ge U>L$ satisfy
the bound automatically; Lemma~\ref{lem:bang} extends it to every
amplitude-limited laboratory pulse.
\end{proof}

\subsection{\texorpdfstring{Strengthening the bound through dynamical constraints}{Strengthening the bound through dynamical constraints}}
\label{sec:slack_extension}

\begingroup
Let $L$ denote the bound of Theorem~\ref{thm:split_exact}.
For every competitive CZ trajectory with $C[u]<U$, the certificate decomposes
the gap between its exposure and $L$ exactly:
\begin{equation*}
C[u]-L=S_{\mathrm{occ}}+S_{\mathrm{int}}+S_{\mathrm{term}},
\end{equation*}
with all three terms nonnegative. The occupation slack $S_{\mathrm{occ}}$ is the
integral of $c-L_jV_j$ over the cells the trajectory traverses; the interface
slack $S_{\mathrm{int}}$ is the sum of the jumps
$V_j(1,\xv)-V_{j+1}(0,\xv)$ across the cell boundaries it crosses; and the terminal
slack $S_{\mathrm{term}}$ is $V_j(\xi,\xv)+n$ evaluated at the pulse endpoint in
its final cell.

Theorem~\ref{thm:split_exact} uses only the nonnegativity of these terms.
Approaching $L$ would require the total slack to become arbitrarily small.
We rule this out by applying additional dynamical reachability bounds to the
same order-$7$ certificate, certifying a strictly positive lower bound on the
total slack of every competing CZ trajectory.
\endgroup

\begin{theorem}[Slack-strengthened certified lower bound]
\label{thm:slack_strengthened}
Under the same assumptions as Theorem~\ref{thm:split_exact},
the stronger bound
\begin{equation}
C_* \;\ge\; B:=2.931287735220
\label{eq:slack_strengthened_bound}
\end{equation}
holds. The decimal value is exact.
\end{theorem}
\begingroup
\begin{proof}
Write $C_{\mathrm{tot}}=C[u]$.
Suppose, for contradiction, that some competitive trajectory achieves
$C_{\mathrm{tot}}<B$. Theorem~\ref{thm:split_exact} then gives
$L\le C_{\mathrm{tot}}<B<U$. The exact slack decomposition then
requires total slack $C_{\mathrm{tot}}-L<B-L$. We show that the dynamical
constraints instead force this slack to exceed $B-L$, contradicting the
assumption $C_{\mathrm{tot}}<B$.

{The amplitude combination}
\begin{equation*}
D=g_1r_2-\sqrt{2}g_2r_1
\end{equation*}
measures the unnormalized mismatch between the two sectors'
excitation ratios ($D=g_1g_2w$, where $w=r_2/g_2-\sqrt{2}r_1/g_1$ is the
synchronization error driving the nonlocal phase in
Appendix~\ref{app:coordinates}) and
obeys the speed limit $|D'|\le c/\sqrt{2}$ under the shared drive, where
$c=|r_1|^2/2+|r_2|^2/4$ is the running cost. Since $r_1=r_2=0$ at both
endpoints, $D(0)=D(S)=0$. An intermediate excursion of $D$ therefore requires
exposure both to reach it and to return to zero. Writing $C(s)$ for the
accumulated exposure, integration from either endpoint gives
\begin{equation*}
|D(s)|\le\frac{\min\{C(s),C_{\mathrm{tot}}-C(s)\}}{\sqrt{2}}.
\end{equation*}
For populations $p_1=|r_1|^2$ and $p_2=|r_2|^2$, the reverse triangle
inequality in Appendix~\ref{app:slack_extension}, Eq.~\eqref{eq:dark_reverse},
supplies a lower bound on $|D|$. A population pair is therefore excluded
whenever this required magnitude exceeds the exposure budget above.
The same appendix uses population speed limits to bound the exposure needed to
reach a population pair from zero excitation. The resulting lower bound also
applies to the return to zero excitation. A pair is excluded if this bound
exceeds either the accumulated exposure $C(s)$ or the remaining exposure
$C_{\mathrm{tot}}-C(s)$.

To strengthen the lower bound, we need to quantify the slack
that every candidate trajectory must incur. We obtain population-dependent
lower bounds on the occupation and interface slacks from the existing
certificate. At each exposure interval or interface, we verify these bounds
for all population pairs except those proved unreachable by the dynamical
constraints. The bounds therefore apply to every candidate trajectory.
Appendix~\ref{app:slack_extension} describes how the bounds are extracted
from the Gram blocks without double counting.

We now turn the local slack estimates into lower bounds on the
accumulated slacks $S_{\mathrm{occ}}$ and $S_{\mathrm{int}}$. We use only contributions
from exposure intervals and interfaces that every candidate trajectory must
traverse. The accumulated exposure $C(s)$ is continuous and nondecreasing from
zero to $C_{\mathrm{tot}}\ge L$, so every candidate fully traverses the selected
intervals below $L$. All internal interfaces also occur at exposure values below
$L$. Integrating the occupation estimates per unit exposure over these intervals
gives a lower bound on $S_{\mathrm{occ}}$; summing the interface estimates at the
crossings gives a lower bound on $S_{\mathrm{int}}$.

Every candidate terminates in the fifth cell, where the terminal contribution
is a polynomial in $C_{\mathrm{tot}}$.
Under the contradiction hypothesis, the normalized excess exposure
\begin{equation*}
\eta=\frac{C_{\mathrm{tot}}-L}{B-L}
\end{equation*}
lies in $[0,1)$. The endpoint polynomial $P_{\mathrm{new}}(\eta)$, defined in
Appendix~\ref{app:slack_extension}, Eq.~\eqref{eq:endpoint_compose}, is the sum
of the certified occupation and interface lower bounds and the terminal
contribution, minus $B-L$. Its positivity therefore establishes that the
unavoidable total slack exceeds the proposed improvement.

All $14$ coefficients of $P_{\mathrm{new}}$ in the degree-$13$ Bernstein basis
\begin{equation*}
B_k^{13}(\eta)=\binom{13}{k}\eta^k(1-\eta)^{13-k},
\qquad k=0,\ldots,13,
\end{equation*}
are strictly positive in exact rational arithmetic. These basis polynomials
are nonnegative and sum to one, so $P_{\mathrm{new}}(\eta)>0$ throughout $[0,1]$.
The exact slack decomposition consequently gives
\begin{equation*}
C_{\mathrm{tot}}-L\ge(B-L)+P_{\mathrm{new}}(\eta)>B-L,
\end{equation*}
contradicting $C_{\mathrm{tot}}<B$. Hence no admissible CZ trajectory
has exposure below $B$, and taking the infimum gives $C_*\ge B$. Excluding
costs below $B$ does not by itself establish a uniform positive gap above $B$.
\end{proof}

The algebraic coefficient identities are verified exactly in
$\mathbb{Q}(\sqrt{2})$, represented by pairs of rational numbers, and the
Bernstein coefficients of $P_{\mathrm{new}}$ are verified strictly positive in
$\mathbb{Q}$, independently of solver precision or numerical sampling.
\endgroup

\section{\texorpdfstring{Certified upper bound and a near-optimal pulse}{Certified upper bound and a near-optimal pulse}}
\label{sec:pulse}

The certified lower bound applies to every admissible CZ gate.
An existence proof is still required to turn a numerical candidate into an
unconditional upper bound.
Small terminal residuals in a floating-point calculation do not by themselves
prove that an exact CZ gate exists.
We therefore use interval arithmetic~\cite{Johansson17} to certify
that an exact CZ gate exists with the exposure bound in
Eq.~\eqref{eq:upper_bound} (Sec.~\ref{sec:interval_closure}).
We inspect a separate compact waveform in the same
low-exposure regime to illustrate phase accumulation and population return
(Sec.~\ref{sec:exposure_dynamics}).

\subsection{Interval-certified gate closure}
\label{sec:interval_closure}
To establish the upper bound, we prove that an exact CZ pulse exists
in the neighborhood of a low-exposure numerical candidate. This candidate is a
piecewise-constant phase pulse of equal-duration segments at total active
duration $S\approx7.74$. Its discretization is specified in Appendix~\ref{sec:methods}.
Exact CZ gate closure under Eq.~\eqref{eq:CZ_cond} requires complete population
return in both sectors, $r_1(S)=r_2(S)=0$, together with the entangling
conditional phase relation $\Phi_2(S)-2\Phi_1(S)\equiv\pi\pmod{2\pi}$.
Vanishing of the real and imaginary parts of the two complex terminal Rydberg
amplitudes supplies four real endpoint conditions, while the conditional phase
constraint provides a fifth. For local certification, we vary the active duration
$S$ and four selected segment drive phases while holding all remaining segment
phases fixed.

We refine these five parameters numerically to high precision.
We then construct a parameter box consisting of small
intervals around the refined duration and four selected phase values, whose
combinations define a set of nearby pulses. Interval arithmetic bounds all
evaluation and roundoff errors when computing the pulse evolution across
the segments. Using these rigorous enclosures, the Krawczyk test~\cite{Krawczyk69,Johansson17}
proves that an exact solution to the five endpoint equations exists within
these ranges. The fifth root-finding equation requires
$\operatorname{Im}(g_2\overline{g_1}^2)=0$, allowing conditional phases of
$0$ or $\pi\pmod{2\pi}$. We therefore also verify with interval arithmetic that
$\operatorname{Re}(g_2\overline{g_1}^2)<0$ across the parameter box, ensuring that
the certified solution realizes the entangling conditional phase $\pi$.

\begingroup
\displaywidowpenalty=10000
Evaluating the Rydberg exposure with interval arithmetic over the same
box yields a bound that holds for every pulse in the family, and therefore
bounds the exact CZ pulse known to be inside. Because the upper endpoint of this
enclosure is strictly below $C_{\mathrm{up}}$, this certifies
\begin{equation}
C_* < C_{\mathrm{up}} := 2.935875225844105.
\label{eq:upper_bound}
\end{equation}
The selected phase indices, residual equations, precision settings, and
Krawczyk contraction bounds are given in Appendix~\ref{sec:methods}.
\par\endgroup

\subsection{\texorpdfstring{A near-optimal pulse and the duration--exposure trade-off}{A near-optimal pulse and the duration--exposure trade-off}}
\label{sec:exposure_dynamics}
\begin{figure}[!t]
\centering
\includegraphics[width=\columnwidth]{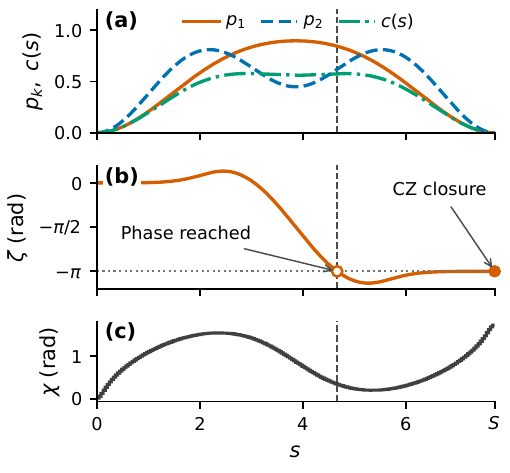}
\caption{The illustrative pulse reaches the target nonlocal
phase before completing population return; about $35\%$ of its exposure
accumulates after that first crossing. This $192$-segment numerical waveform
has duration $S=7.725$ and exposure $C\approx2.936$. (a)~Rydberg
populations $p_k=|r_k|^2$ for the $\bket{01}$ and $\bket{10}$ inputs
(solid) and the $\bket{11}$ input (dashed), together with the exposure rate $c(s)=(2p_1+p_2)/4$
(dash-dotted), whose integral is $C$.
(b)~The continuous nonlocal phase $\zeta(s)=\arg g_2-2\arg g_1$, with target
$-\pi$ modulo $2\pi$. (c)~The actual piecewise-constant laser phase $\chi(s)$.
The dashed vertical marker denotes the first downward
crossing of $\zeta=-\pi$; substantial Rydberg populations
remain then. Population return and
the target phase coincide only at the terminal point of the plotted waveform.
Active time is $s=\Omax t$ for this maximum-amplitude pulse. This numerical
waveform illustrates the gate dynamics; the upper bound is established using
the separate interval-certified pulse described in Sec.~\ref{sec:interval_closure}.}
\label{fig:pulse}
\end{figure}

We use a pulse close to the certified exposure floor to illustrate
the two requirements for CZ closure: the nonlocal phase must reach $\pi$
modulo $2\pi$, and both sectors must return to the computational manifold.
Figure~\ref{fig:pulse} shows an optimized waveform in the same low-exposure
regime as Locher \emph{et al.}~\cite{Locher25}.

In Fig.~\ref{fig:pulse}(a), the Rydberg population $p_1=|r_1|^2$
in the $\bket{01}$ and $\bket{10}$ sectors forms a broad central plateau,
while $p_2=|r_2|^2$ in the $\bket{11}$ sector reaches a local minimum at the
pulse midpoint. This dip reduces the $\bket{11}$ sector's contribution to
Rydberg exposure while the $\bket{01}$ and $\bket{10}$ sectors continue
accumulating the ground-state phase $\Phi_1=\arg g_1$. Both Rydberg populations then return to zero within numerical precision at $s=S$.

In this low-exposure waveform, about $35\%$ of the total exposure accumulates after the nonlocal phase first reaches $-\pi$.
The continuous nonlocal phase $\zeta(s):=\arg g_2(s)-2\arg g_1(s)$
is shown in Fig.~\ref{fig:pulse}(b) (see Appendix~\ref{app:coordinates},
Lemma~\ref{lem:ratio}). At this first crossing, both populations are still substantial.
The phase then overshoots before returning to $-\pi$ at $s=S$,
when both populations also return to zero within numerical precision.

We next examine how exposure varies with gate duration by locally optimizing
pulses at fixed durations (Appendix~\ref{sec:methods}). Compared with the
time-optimized candidate near the numerical speed limit ($S\approx7.612$)~\cite{JP22,Fromonteil24},
slightly longer pulses have lower exposure. The scan finds a nearly flat minimum
over $S\in[7.7,8.5]$, with less than $0.1\%$ variation across this interval.
At longer durations, the exposure of the same local-search candidates increases.

\begingroup
Compared with the numerical speed-limit candidate, the illustrative waveform
reduces exposure by about $0.7\%$ at a roughly $1.5\%$ longer duration.
To assess whether this exposure reduction outweighs the cost of a longer gate,
we use the simple error model $\Gamma\TR+\gamma T$, combining decay
loss~\cite{JP22,Locher25} with a technical-error contribution proportional to
duration. Here $\gamma$ is the effective coefficient of the technical-error term.
For these two candidate pulses, the longer pulse is preferable when
$\gamma/\Gamma$ is below approximately $0.19$; the numerical comparison is
given in Appendix~\ref{sec:methods}.
\par\endgroup

\section{Discussion and outlook}
\label{sec:discussion}

These results establish that a nonzero floor on the Rydberg exposure
persists for globally driven CZ gates even when the gate duration is left unrestricted.
The certified lower and upper bounds confine $C_*$ to the window
\begin{equation}
2.9312 \le C_* < 2.9359,
\label{eq:certified_window_disc}
\end{equation}
with a relative gap below $0.16\%$ before rounding.
This narrow window leaves little room for further reductions in decay loss within the present model.
Decay loss is only one contribution to the total gate error.
Duration-dependent technical errors must also be considered when choosing the gate duration,
as illustrated by the comparison in Sec.~\ref{sec:exposure_dynamics}.

Inside the present shared-drive model, a natural question is whether the remaining
gap reflects a fundamental limitation of the moment-SOS relaxation.
{We adapt the occupation-measure framework of Lasserre \emph{et al.}~\cite{Lasserre08}
by using measures that record accumulated Rydberg exposure in state--control regions.
With this formulation, we establish asymptotic completeness of the full real-coefficient
SOS dual hierarchy in Appendix~\ref{app:sos-convergence}: for any fixed exposure partition
as in Sec.~\ref{sec:exposure_partition}, the optimal values $L_d$ of the order-$d$
SOS dual programs satisfy $L_d\uparrow C_*$
as $d\to\infty$.} The model-specific proof rigorously overcomes the absence of a uniform
terminal-time bound and the vanishing of the running cost on the computational manifold.
While the unsplit hierarchy converges in the infinite-degree limit, driving $d$
higher is computationally prohibitive in practice.
This convergence result does not guarantee that a given finite
relaxation order will close the remaining gap.
Partitioning cumulative exposure enables exact algebraic certification of $C_*\ge2.93$
already at order~$6$. At finite relaxation order, adaptive partition grids and higher
algebraic orders provide a systematic path to narrowing the remaining window further.

Lowering the dimensionless exposure below the certified bound requires leaving
the shared-drive, perfect-blockade model of Sec.~\ref{sec:model}.
Dynamically, the floor arises because the shared field drives the
$\bket{01}$ and $\bket{10}$ sectors with Rabi frequency $\Omega$ and the
$\bket{11}$ sector with $\sqrt{2}\Omega$; relaxing the dynamics to allow
independent control over the two sectors removes this lock and drives the exposure
infimum to zero (Theorem~\ref{thm:relax}).
In physical architectures, independently addressing each atom with local
laser beams does not decouple the two sectors in this idealized way, and whether
continuous pulse shaping with site-resolved fields can undercut the certified bound remains
an open question. Further directions include auxiliary dressing, dual-species control,
and operation at finite blockade $B\sim\Omax$~\cite{DDP25}.
The narrow certified window motivates seeking substantial reductions in dimensionless
exposure through changes in control resources or the physical model.

\begin{acknowledgments}
This work was partially supported by the National Key R\&D Program
of China under Grant No.~2023YFA1009401, the National Natural Science
Foundation of China under Grant Nos.~12404568, 12171324, 92576114, 12447107, and the Guangdong Provincial Quantum Science Strategic Initiative (Grant No. GDZX2403008, GDZX2503001).
\end{acknowledgments}

\section*{Code and data availability}
The data and code that support the findings of this article are openly
available at \url{https://github.com/wanda0929/Lowerbound_pub}, with
version~1.1 archived on Zenodo~\cite{repo_pub}. The deposit includes the numerical pulse waveforms, the
SDP formulations, the exact $\mathbb{Q}(\sqrt{2})$ dual certificates,
and the automated verification suites. It also includes Lean formalization files,
with documentation of their verification scope and assumptions.
Reproduction instructions are given with that
deposit; the numerical methods and the certificate specifications are in
Appendix~\ref{sec:methods}.

\clearpage
\onecolumngrid
\appendix

\section{No-go theorems for shortcut relaxations}
\label{sec:nogo}

This appendix proves two facts used in the main text. An isolated two-level atom can
acquire any ground-state phase with exposure driven to zero. If the two driven cases
also have independent controls, an exact CZ has the same vanishing infimum. Neither
statement bounds the shared-drive $C_*$. Together they say that a strictly positive floor,
if it exists, has to come from one laser locking both sectors, not from a per-sector phase
condition and not from unlocking the drive.

The first fact is the following.
\begin{theorem}[A single two-level phase is free]
\label{thm:free}
For an uncoupled two-level system ($k=1$ or $2$) in active time with $|u|=1$,
unrestricted duration, and unrestricted drive-phase modulation, the infimum of the
exposure required to accumulate an arbitrary ground phase $\Phi$ with exact loop
closure $r(S)=0$ is zero.
\end{theorem}
\begin{proof}
We construct a family of constant-detuning pulses that reach the same target phase
with exposure tending to zero. In units $\Omax=1$ with coupling $c=\sqrt{k}/2$, take the
constant-detuning drive $\chi(s)=\Delta s$ ($|u|=1$). In the rotating frame $g=\tilde
 g\,\ee^{-\ii\chi/2}$, $r=\tilde r\,\ee^{\ii\chi/2}$, the dynamics~\eqref{eq:schrod} is
autonomous with generator $H_{\mathrm{eff}}=\begin{pmatrix}-\Delta/2 & c\\ c &
\Delta/2\end{pmatrix}$ and frequency $\omega=\sqrt{c^2+\Delta^2/4}$:
\begin{equation}
\tilde g(s)=\cos\omega s+\ii\frac{\Delta}{2\omega}\sin\omega s,
\qquad
\tilde r(s)=-\ii\,\frac{c}{\omega}\sin\omega s .
\end{equation}
Let $m=1,2,\ldots$ count complete oscillations of the Rydberg population, whose
period is $\pi/\omega$. At the $m$th return, $S_m=m\pi/\omega$, the Rydberg amplitude
closes exactly, $r(S_m)=0$, and $g(S_m)=(-1)^m\ee^{-\ii\Delta S_m/2}$, so the acquired
phase is $\Phi_m=m\pi\bigl(1-\Delta/(2\omega)\bigr)\pmod{2\pi}$. The exposure is
\begin{equation}
E_m=\frac{c^2}{\omega^2}\int_0^{S_m}\sin^2\omega s\,\dd s=\frac{m\pi c^2}{2\omega^3}.
\end{equation}

The trivial target $\Phi=0\pmod{2\pi}$ is attained with zero exposure at $S=0$.
For a nontrivial target, choose its representative $\Phi\in(0,2\pi)$ and set
$Q=\Phi/\pi$. We now choose the detuning separately for each $m$, keeping the target
phase fixed. Define $\omega_m=\sqrt{c^2+\Delta_m^2/4}$ and require
$1-\Delta_m/(2\omega_m)=Q/m$, so that $\Phi_m=\Phi\pmod{2\pi}$.
Such a $\Delta_m$ exists uniquely because the map
$\Delta\mapsto1-\Delta/(2\omega)$ is continuous and strictly monotone with range $(0,2)$.
Along this family,
\[
\omega_m^2=\frac{c^2}{2Q/m-Q^2/m^2}=\Theta(m),
\]
so $\omega_m=\Theta(m^{1/2})$. The exposure then falls as
\begin{equation}
E_m=\frac{m\pi c^2}{2\omega_m^3}=\Theta(m^{-1/2})\xrightarrow[m\to\infty]{}0 .
\end{equation}
For a nontrivial target the infimum is zero but is not attained: $E=0$ would require
$r\equiv0$ almost everywhere, which freezes the phase. The required duration diverges,
$S_m=\Theta(m^{1/2})$, which the unrestricted-duration problem permits.
\end{proof}

The second fact is the following.
\begin{theorem}[An unlocked CZ has exposure infimum zero]
\label{thm:relax}
If the $\bket{01}$ and $\bket{10}$ sectors share one bounded control
and the $\bket{11}$ sector has an independent bounded control, the infimum
of the exposure of an exact CZ is zero.
\end{theorem}
\begin{proof}
Give the $\bket{11}$ sector its own drive $v_2$ and leave the
$\bket{01}$ and $\bket{10}$ sectors idle, $v_1\equiv 0$. Theorem~\ref{thm:free} then yields $\Phi_2=\pi$ with $r_2(S)=0$ and
$E_2\to 0$, while $\Phi_1=0$ and $E_1=0$. The CZ condition $\Phi_2-2\Phi_1=\pi$ holds and
the average exposure $(2E_1+E_2)/4\to 0$.
\end{proof}

\section{Analytical lower bound via a progress coordinate}
\label{sec:analytical}

To obtain an analytical lower bound on $C=\int_0^S c(\xv(s))\,\dd s$, we compare
the blockaded collective state $A$ with a companion state $B$ describing two
noninteracting atoms under the same drive. Both states are expressed in the basis
$\{\bket{11},\bket{W},\bket{rr}\}$.
Their overlap $z:=A^\dagger B$ equals $+1$ at the start and $-1$ at every CZ endpoint,
so its real part $q=\operatorname{Re}z$ must change from $+1$ to $-1$.
We then relate $|q'|$ to the Rydberg populations to bound the exposure required for
this change.

In that basis the third entry of $A$ vanishes, because perfect blockade forbids
$\bket{rr}$. The companion $B$ is the symmetric square of $(g_1,r_1)$, so $\bket{rr}$ may
be occupied, and $|g_1|^2+|r_1|^2=1$ implies $\|B\|=1$:
\begin{equation}
A(s) \;=\; \begin{pmatrix} g_2 \\ r_2 \\ 0 \end{pmatrix},
\qquad
B(s) \;=\; \begin{pmatrix} g_1^2 \\ \sqrt{2}g_1 r_1 \\ r_1^2 \end{pmatrix}.
\label{eq:AB_def}
\end{equation}
The laser couples $\bket{11}$ to $\bket{W}$ through the shared Hamiltonian $H_0$, which
both $A$ and $B$ feel and which never reaches $\bket{rr}$ (its third row and column
vanish). Blockade removes the additional $\bket{W}\leftrightarrow\bket{rr}$ coupling $H_1$
from the physical collective motion of $A$, while the companion $B$ still evolves under
$H_0+H_1$ because it is constructed as if blockade were absent:
\begin{equation}
\begin{gathered}
H_0 = \frac{1}{\sqrt{2}}\begin{pmatrix} 0 & u & 0 \\ u^* & 0 & 0 \\ 0 & 0 & 0
\end{pmatrix},
\qquad
H_1 = \frac{1}{\sqrt{2}}\begin{pmatrix} 0 & 0 & 0 \\ 0 & 0 & u \\ 0 & u^* & 0
\end{pmatrix},
\\
\ii A' = H_0 A,
\qquad
\ii B' = (H_0 + H_1) B.
\end{gathered}
\label{eq:H0_H1}
\end{equation}
The components in~\eqref{eq:AB_def} give the explicit overlap $z=\bar{g}_2
g_1^2+\sqrt{2}\bar{r}_2 g_1 r_1$. Substituting the endpoints already used above recovers
$z(0)=1$ and $z(S)=-1$. Because $A$ and $B$ share $H_0$, that generator cancels in $z'$,
and only the withheld $H_1$ remains:
\begin{equation}
z' \;=\; -\ii A^\dagger H_1 B \;=\; -\frac{\ii}{\sqrt{2}}\bar{r}_2 u r_1^2.
\label{eq:z_deriv}
\end{equation}
The identity for $z'$ then limits how fast $q$ can move:
\begin{equation}
|q'| \;\le\; |z'| \;\le\; \frac{x\sqrt{y}}{\sqrt{2}},
\qquad x = |r_1|^2,\quad y = |r_2|^2.
\label{eq:q_dot_bound}
\end{equation}
The running cost is made of the same populations, $c=(2x+y)/4$.
The ratio $c/|q'|$ is therefore the cost per unit overlap progress. If the populations
vanish, $q$ freezes; finishing the trip from $+1$ to $-1$ requires exposure.

For $0\le x,y\le1$, the inequality $3x\sqrt y\le2x+y$ and
Eq.~\eqref{eq:q_dot_bound} give $|q'|\le(2\sqrt2/3)c$.
Since $q(0)=1$ and $q(S)=-1$, integration yields
$C\ge3/\sqrt2\approx2.12132$ for every exact CZ pulse, regardless of its duration.

For a stronger bound, we use the additional constraint
$|z|\le\Gamma(x,y):=(1-x)\sqrt{1-y}+\sqrt{2x(1-x)y}$,
which follows from the two terms of $z=A^\dagger B$. At
overlap level $|q|=\xi$, only those populations with $\Gamma(x,y)\ge\xi$ can occur.

\begin{theorem}[Companion-state overlap bound]
\label{thm:fiber}
Every exact CZ pulse has exposure $C\ge C_{\mathrm{fib}}$, and consequently
\begin{equation}
C_* \;\ge\; C_{\mathrm{fib}} \;:=\; 2\int_0^1 c_{\mathrm{fib}}(\xi)\,\dd\xi,
\label{eq:C_fib}
\end{equation}
where $c_{\mathrm{fib}}(\xi)$ is the cheapest cost per unit overlap progress at level
$\xi$
\begin{equation}
c_{\mathrm{fib}}(\xi) \;:=\; \inf_{\substack{0 < x, y \le 1 \\ \Gamma(x,y) \ge \xi}}
\frac{\sqrt{2}}{4}\left( \frac{2}{\sqrt{y}} + \frac{\sqrt{y}}{x} \right).
\label{eq:c_fib_def}
\end{equation}
\end{theorem}
\begin{proof}
Write $x=|r_1|^2$, $y=|r_2|^2$, and $c(x,y)=(2x+y)/4$. The triangle bound
$|z|\le\Gamma(x,y)$ already restricts the populations at level $|q|$.
Since $|q'|\le|z'|$ and~\eqref{eq:q_dot_bound} holds, every point with $q'\ne0$ satisfies
\begin{equation}
\frac{c(x,y)}{|q'|}
\;\ge\; \frac{\sqrt{2}}{4}\left(\frac{2}{\sqrt{y}}+\frac{\sqrt{y}}{x}\right).
\label{eq:fiber_pointwise}
\end{equation}
The same $(x,y)$ is feasible for the infimum at level $|q|$, because
$\Gamma(x,y)\ge|z|\ge|q|$ and $q'\ne0$ forces $x,y>0$. Therefore the ratio is at least
$c_{\mathrm{fib}}(|q|)$.
Thus $c(x,y)\ge c_{\mathrm{fib}}(|q|)|q'|$, with the product defined
as zero when $q'=0$. Since $q$ crosses every level from $1$ to $-1$, integration gives
the following inequalities, understood by truncating $c_{\mathrm{fib}}$ at a finite
cutoff and then sending that cutoff to infinity:
\begin{equation}
C=\int_0^S c(x(s),y(s))\,\dd s
\;\ge\;\int_0^S c_{\mathrm{fib}}(|q|)|q'|\,\dd s
\;\ge\;\int_{-1}^{1}c_{\mathrm{fib}}(|\xi|)\,\dd\xi
=2\int_0^1c_{\mathrm{fib}}(\xi)\,\dd\xi.
\end{equation}
\end{proof}

For $0<\xi<1$, the objective in Eq.~\eqref{eq:c_fib_def} strictly
decreases with $x$ at fixed $y$, so its minimum lies on $\Gamma(x,y)=\xi$. The lowest
feasible stationary value on this boundary gives $c_{\mathrm{fib}}(\xi)$; numerical
quadrature yields $C_{\mathrm{fib}}\approx 2.4801$. The remaining slack comes from dropping the phase of the
overlap, applying the triangle bound on $|z|$, and projecting onto a single scalar; the
global certificate of Sec.~\ref{sec:moments_sos} retains the unprojected dynamics.

\section{Mathematical specification of the exact five-cell certificate}
\label{app:split_certificate}

This appendix specifies the exact five-cell certificates and their independent verification.
Appendix~\ref{app:terminal_quotient} defines the invariant polynomial representation
and reduces the terminal inequality on the CZ orbit to a univariate condition.
Appendix~\ref{subsec:positivity_cones} gives SOS representations of the drift,
interface, and terminal inequalities.
Appendix~\ref{app:auditor} describes the exact checks of polynomial identities and
Gram-matrix positive semidefiniteness over $\mathbb Q(\sqrt2)$.
Appendix~\ref{app:order7_certificate} presents the order-$7$ certificate and its exact verification.
Appendix~\ref{app:slack_extension} strengthens its lower bound using dynamical reachability constraints and certified slack estimates.

\subsection{Invariant quotient and the CZ boundary functional}
\label{app:terminal_quotient}
Normalization and the phase symmetries of Sec.~\ref{sec:sos_reductions}
simplify the polynomial representation of the three inequalities in~\eqref{eq:split_pop}.
At a CZ endpoint, the nonlocal phase is fixed to $\pi$, while the remaining common
single-qubit phase can be corrected by local $Z$ rotations. Invariant state
polynomials are independent of this remaining phase, so the terminal inequality
reduces to a univariate condition in the exposure coordinate $\xi$.

Let $\mathbb F=\mathbb Q(\sqrt2)$ and $\zv=(u,\xv)$. On each
exposure cell we work in the quotient algebra closed under complex conjugation
\begin{equation}
\mathcal A
=\mathbb F(\ii)[\xi,\bar{\zv},\zv]/\mathcal I,
\qquad
\mathcal I=\left\langle
\bar uu-1,\ \bar g_1g_1+\bar r_1r_1-1,\
\bar g_2g_2+\bar r_2r_2-1
\right\rangle.
\label{eq:quotient_algebra}
\end{equation}
A mixed monomial and its total degree are
\begin{equation}
m_{p,\alpha,\beta}(\xi,\zv)=\xi^p\bar{\zv}^\alpha \zv^\beta,
\qquad
\deg m_{p,\alpha,\beta}=p+|\alpha|+|\beta|,
\label{eq:mixed_monomial}
\end{equation}
where $p \in \mathbb{N}_0$ is the cell coordinate exponent, and
$\alpha=(\alpha_1,\ldots,\alpha_5)$ and
$\beta=(\beta_1,\ldots,\beta_5)$ are exponent multi-indices in
$\mathbb{N}_0^5$ for the ordered base variables $\zv=(u,g_1,r_1,g_2,r_2)$.
For $\ell=1,\ldots,5$, the component $\alpha_\ell$ is the exponent of
$\bar{\zv}_\ell$, while $\beta_\ell$ is the exponent of $\zv_\ell$:
\[
\bar{\zv}^{\alpha}\zv^{\beta}
=\prod_{\ell=1}^5 \bar{\zv}_\ell^{\alpha_\ell}\zv_\ell^{\beta_\ell}.
\]
We write $|\alpha|=\sum_{\ell=1}^5\alpha_\ell$ and
$|\beta|=\sum_{\ell=1}^5\beta_\ell$.

The two charges are those of the {optical-phase} and {qubit-phase}
symmetries of Sec.~\ref{sec:sos_reductions}:
\begin{equation}
q(u)=(1,0),\quad q(g_1)=(0,1),\quad q(r_1)=(-1,1),
\quad q(g_2)=(0,2),\quad q(r_2)=(-1,2).
\label{eq:variable_charges}
\end{equation}
The exposure coordinate $\xi$ is unchanged by both rotations and has
charge $(0,0)$. For an exponent multi-index $\alpha$, write
$q(\alpha)=\sum_{\ell=1}^5\alpha_\ell q(\zv_\ell)$.
The total charge of the mixed monomial in Eq.~\eqref{eq:mixed_monomial} is
\begin{equation}
\begin{aligned}
q(m_{p,\alpha,\beta})
&=\sum_{\ell=1}^5(\beta_\ell-\alpha_\ell)q(\zv_\ell)\\
&=q(\beta)-q(\alpha).
\end{aligned}
\label{eq:monomial_total_charge}
\end{equation}
The minus sign accounts for the conjugated variables. For the coefficient
representation of invariant scalar polynomials, we retain precisely the
monomials with total charge $(0,0)$, or equivalently
$q(\alpha)=q(\beta)$. Monomials inside an SOS factor may have nonzero
charge; their conjugate products are invariant whenever the two
monomials have the same charge.

To see how this condition reduces a Gram matrix, let
$b=(b_1,\ldots,b_N)^{\mathsf T}$ be a chosen reduced monomial basis, with
$q_i=q(b_i)$, and suppose $b^*Qb$ represents an invariant polynomial in
the quotient algebra~\eqref{eq:quotient_algebra}. Write
$\boldsymbol\phi=(\phi_1,\phi_2)$ for the optical and qubit rotation angles.
Under a rotation, $\bar b_i b_j$ acquires the phase
$\ee^{\ii(q_j-q_i)\cdot\boldsymbol\phi}$. The uniformly averaged Gram
matrix therefore has entries
\begin{equation}
\begin{aligned}
\widehat Q_{ij}
&=\frac{Q_{ij}}{(2\pi)^2}\int_{[0,2\pi]^2}
\ee^{\ii(q_j-q_i)\cdot\boldsymbol\phi}\,\dd^2\boldsymbol\phi\\
&=\begin{cases}
Q_{ij},&q_i=q_j,\\
0,&q_i\ne q_j.
\end{cases}
\end{aligned}
\label{eq:charge_gram_average}
\end{equation}
Each rotated Gram matrix is a unitary diagonal conjugation of $Q$, so
its positive semidefiniteness is preserved by the average. The represented
polynomial is unchanged because it is invariant. After grouping the basis
by charge, let $b_q$ collect the monomials of charge $q$ and let $Q_q$ be
the corresponding principal block. Then
\begin{equation}
b^*\widehat Qb=\sum_q b_q^*Q_qb_q,
\qquad
\widehat Q\succeq0\ \Longleftrightarrow\ Q_q\succeq0\ \text{for every }q.
\label{eq:charge_gram_blocks}
\end{equation}
The matrix-positivity constraint splits into smaller constraints, while
the polynomial coefficient equations can still couple different blocks.

\begin{figure*}[t]
\centering
\includegraphics[width=\textwidth]{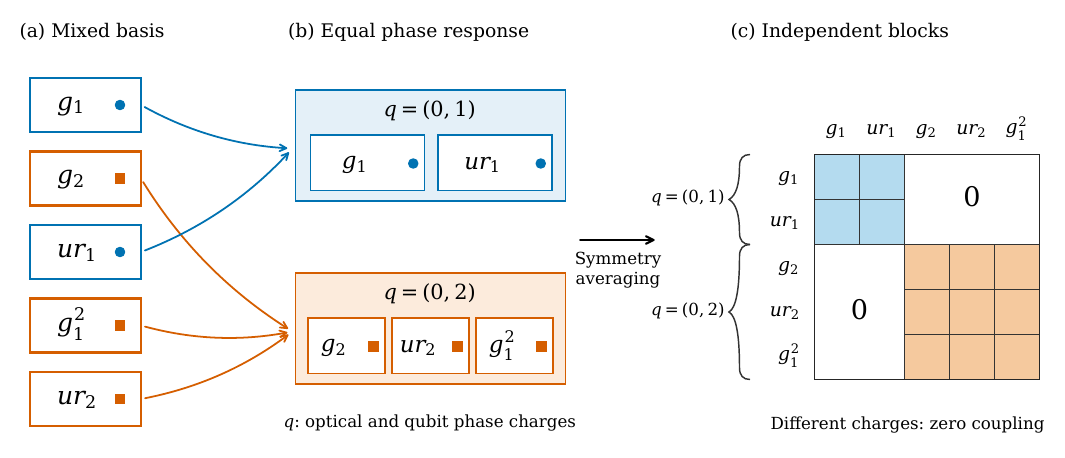}
\caption{Averaging over the optical and qubit phase rotations
eliminates couplings between different phase-response classes, yielding
independent matrix blocks. A monomial's charge $q$ is the integer pair
specifying its response to these two rotations. (a,b)~The illustrative basis
groups $g_1$ and $ur_1$ at charge $(0,1)$, and $g_2$, $ur_2$, and $g_1^2$
at charge $(0,2)$. (c)~In a Gram representation, the polynomial is expressed
as conjugate products of basis monomials weighted by matrix entries.
Symmetry averaging sets cross-charge entries to zero while preserving the
invariant polynomial and positive semidefiniteness; reordering the basis
alone does not remove these entries. Colored cells denote allowed couplings,
not numerical matrix values. The displayed $2\times2$ and $3\times3$ blocks
illustrate grouping before {face reduction}; they are not the
compressed certificate basis. For the full unsplit order-$6$ drift Gram,
charge grouping and {face reduction}, together with reflection
symmetry and basis rephasing, give $113$ real PSD blocks,
the largest $75\times75$. These dimensions do not describe the split
order-$7$ program.}
\label{fig:method}
\end{figure*}

For example, the five basis monomials in Fig.~\ref{fig:method} form the
two groups
\[
\begin{aligned}
b_{(0,1)}&=(g_1,ur_1)^{\mathsf T},\\
b_{(0,2)}&=(g_2,ur_2,g_1^2)^{\mathsf T}.
\end{aligned}
\]
Thus their $5\times5$ Gram matrix reduces to a $2\times2$ block and a
$3\times3$ block. Within the first block, $\bar g_1ur_1$ has total charge
$(0,0)$ and survives the average; the cross-block product $\bar g_1g_2$
has charge $(0,1)$ and averages to zero. This example illustrates charge
grouping before the separate zero-cost-face compression described in
Appendix~\ref{subsec:positivity_cones}.

The normal form first removes every common $\bar uu$ factor. For each $k$, it then removes
$m_k=\min(\alpha_{g_k},\beta_{g_k})$ common $\bar g_kg_k$ factors and expands
$(1-\bar r_kr_k)^{m_k}$ exactly. A normal monomial therefore contains no
simultaneous barred and unbarred power of $u$, $g_1$, or $g_2$. From each
conjugate pair we retain the lexicographically earlier exponent tuple in the
variable order displayed above. Reflection rephasing rescales basis
representatives by $\ii$
precisely when the total degree $k$ in $u$ and $\bar{u}$ is odd. Conjugation of the
amplitudes together with $u\mapsto-\bar u$ sends that monomial to $(-1)^k$ times
its conjugate, so the invariant expansion coefficients are purely imaginary for odd $k$
and real for even $k$. The factor of $\ii$ absorbs the sign for odd $k$, ensuring that
every Gram block is real symmetric while strictly preserving polynomial equivalence
and positive semidefiniteness.

The CZ terminal set~\eqref{eq:cz_terminal_set} is a one-parameter orbit.
After charge invariance, the restriction
\begin{equation}
(\tau p)(\xi):=p(\xi,1,0,-1,0)
\label{eq:tau_main}
\end{equation}
evaluates every invariant state polynomial at a single representative and
yields one univariate polynomial in $\xi$. On monomials,
\begin{equation}
\tau m_{p,\alpha,\beta}
=
\begin{cases}
0,
&\alpha_{r_1}+\beta_{r_1}+\alpha_{r_2}+\beta_{r_2}>0,\\[2mm]
(-1)^{\alpha_{g_2}+\beta_{g_2}}\xi^p,
&\text{otherwise}.
\end{cases}
\label{eq:terminal_functional}
\end{equation}
Every $\xv\in\mathcal X_{\mathrm{CZ}}$ has the form
$(g_1,r_1,g_2,r_2)=(\ee^{\ii\theta},0,-\ee^{2\ii\theta},0)$. Any Rydberg factor
vanishes; otherwise charge invariance cancels the $\theta$ dependence and
leaves the sign in~\eqref{eq:terminal_functional}. Thus
$\tau p=p(\xi,\xv)$ for every invariant $p$ and every $\xv\in\mathcal X_{\mathrm{CZ}}$,
so the terminal inequality $V_j+n\ge0$ on
$[0,1]\times\mathcal{X}_{\mathrm{CZ}}$ is the univariate condition
$\tau V_j+n\ge0$.

\subsection{The three positivity cones}
\label{subsec:positivity_cones}
To obtain an exposure bound valid for every admissible CZ pulse, we certify
the three inequalities in~\eqref{eq:split_pop} throughout their respective domains.
SOS representations reduce this verification to polynomial identities
and PSD Gram matrices.
The drift and interface cones are built in the same quotient
algebra~\eqref{eq:quotient_algebra}. The terminal cone uses the functional
$\tau$ of the previous subsection.

For stationary certificates $W(\mathbf x)$ of Eq.~\eqref{eq:erev-stationary},
which are independent of $\xi$, we reduce the drift Gram matrix dimension
by exploiting the zero-cost set $F=\{r_1=r_2=0\}$.
On $F$, no Rydberg population is present and $c=0$.
A drive-phase shift of $\pi$ sends $u$ to $-u$, so Eq.~\eqref{eq:vector_field}
gives $f_{-u}=-f_u$.
The directional derivative $f_uW=\nabla W\cdot f_u$ is the trajectory
derivative $\dd W/\dd s$, so nonnegativity of $c+f_uW$ for
both controls $u$ and $-u$ forces $f_uW=0$ on $F\times S^1$. From the
stationary identity $c+f_uW\equiv\sum_\ell h_\ell^2\pmod{\mathcal I}$, the sum
of squares $\sum_\ell h_\ell^2$ vanishes identically on $F\times S^1$,
requiring every real factor $h_\ell$ to vanish on $F\times S^1$. Restricting
the factor polynomial basis to elements vanishing on $F\times S^1$ therefore
discards no valid stationary certificate; we refer to this step as {face reduction}.

For each retained charge $q$ (possibly nonzero; one representative of the pair $q,-q$), let
$\mathbf p_{q,d}$ be the SOS factor column of quotient-normal mixed
monomials~\eqref{eq:mixed_monomial} of total degree at most $d$ in that
charge sector. To implement {face reduction}, we construct $\mathbf h_{q,d}$ as a basis
for the polynomials in $\operatorname{span}\mathbf p_{q,d}$ that vanish on $F$.
A monomial containing $r_k$ or $\bar r_k$ already vanishes on $F$.
The remaining monomials are grouped by their restriction character
$(p,\beta_u-\alpha_u,\beta_{g_1}-\alpha_{g_1},\beta_{g_2}-\alpha_{g_2})$;
monomials in the same group agree on $F$. Thus, for an ordered group
$(m_0,\ldots,m_s)$, the combinations
$\sum_{i<\ell}m_i-\ell m_\ell$, $\ell=1,\ldots,s$, vanish on $F$.
These vectors form an exact integer basis for the kernel of
$\mathbf p_{q,d}\mapsto\mathbf p_{q,d}|_{r_1=r_2=0}$.
In the cone formulas below, $\mathbf p_{q,d}$ and
$\mathbf h_{q,d}$ include the reflection rephasing above; the resulting
real representation uses $\mathbb F$ for exact coefficient coordinates
and certified Gram matrices.
The cone that certifies
the drift inequality consists of polynomials
\begin{equation}
\begin{split}
\mathcal Q^{\mathrm{occ}}_d:=\biggl\{&
\sum_q\mathbf h_{q,d}^{*}Q_q\mathbf h_{q,d}
+\xi(1-\xi)\sum_q\mathbf p_{q,d-1}^{*}R_q\mathbf p_{q,d-1}\\
&+\xi(1-\xi)\sum_{\ell}s_\ell|m_\ell|^2:
Q_q\succeq0,\ R_q\succeq0,\ s_\ell\ge0\biggr\}\subset\mathcal A.
\end{split}
\label{eq:occupation_cone}
\end{equation}
{In the first family, every SOS factor vanishes on $F$.}
The second family incorporates the cell constraint through the multiplier
$\xi(1-\xi)$, which is nonnegative on $[0,1]$.
In the third family, $\{m_\ell\}$ ranges over quotient-normal support monomials in the
joint variables $(\xi,\zv)$ of total degree at most $d-1$, retaining exactly one
representative per distinct reduced squared-magnitude polynomial $|m_\ell|^2$.
The last family can be absorbed into the diagonal entries of $R_q$ and
therefore does not enlarge the cone. Every term in~\eqref{eq:occupation_cone}
is nonnegative on $\mathcal K\times[0,1]$.

The interface polynomial $V_j(1,\xv)-V_{j+1}(0,\xv)$ depends only on the state.
We therefore form $\mathbf p^{\mathrm{int}}_{q,d}$ by retaining only the state
monomials in $\mathbf p_{q,d}$, and set
\begin{equation}
\mathcal Q^{\mathrm{int}}_d
:=\left\{\sum_q(\mathbf p^{\mathrm{int}}_{q,d})^*Q_q\mathbf
p^{\mathrm{int}}_{q,d}:Q_q\succeq0\right\}.
\label{eq:interface_cone}
\end{equation}
Since $\tau V_j+n$ depends only on $\xi$, we certify its nonnegativity on
$[0,1]$ using a univariate SOS cone.
With $\mathbf a_d=(1,\xi,\ldots,\xi^d)^\top$ and
$\mathbf a_{d-1}=(1,\xi,\ldots,\xi^{d-1})^\top$, the terminal cone is
\begin{equation}
\mathcal H_d([0,1])
:=\left\{\mathbf a_d^*Q\mathbf a_d
+\xi(1-\xi)\mathbf a_{d-1}^*R\mathbf a_{d-1}:Q,R\succeq0\right\}.
\label{eq:terminal_cone}
\end{equation}
The archived order-$6$ five-cell point satisfies
\begin{subequations}
\label{eq:split_dual_conditions}
\begin{align}
q_j&:=c-L_jV_j\in\mathcal Q^{\mathrm{occ}}_6,
&& j=1,\ldots,5,\\
h_j&:=V_j(1,\cdot)-V_{j+1}(0,\cdot)\in\mathcal Q^{\mathrm{int}}_6,
&& j=1,\ldots,4,\\
b_j&:=\tau V_j+n\in\mathcal H_6([0,1]),
&& j=1,\ldots,5.
\end{align}
\end{subequations}
These three memberships are SOS certificates of the three
inequalities of~\eqref{eq:split_pop}; they supply the nonnegativity
conditions required by the integration argument in~\eqref{eq:split_weak_duality}.
{At order $6$, the first term of Eq.~\eqref{eq:occupation_cone} contains $113$ Gram
blocks $Q_q$ per cell; these are not the $113$ unsplit blocks of maximum side $75$
in Table~\ref{tab:sos_progression}.}

\subsection{The exact auditor}
\label{app:auditor}
Numerical SDP optimization provides a candidate certificate.
To establish a rigorous lower bound, the auditor verifies the archived certificate
independently of the solver, checking its polynomial identities and the positive
semidefiniteness of its Gram matrices.
It takes the order, the rational knots, the exposure ceiling, and the
serialized exact data. It regenerates the monomial basis, the normal-form
reduction, the map $\tau$, and the Gram layout of the three cones, then checks
that each polynomial in~\eqref{eq:split_dual_conditions} equals its
SOS expression coefficient by coefficient. For the
order-$6$ five-cell certificate, this gives $9{,}180$ identities
over $\mathbb F$, with no floating-point tolerance.

It remains to certify the positive semidefiniteness of every real symmetric
Gram matrix over $\mathbb F$. Exact-zero matrices are accepted directly.
For each nonzero matrix, rows and columns that are identically zero in
$\mathbb F$ are removed. The original matrix is PSD if and only if the
remaining principal block $Q$ is PSD. The first route proves positive
definiteness by showing that the factorization error is smaller than a
positive margin $\delta$. It uses floating eigendecomposition and Cholesky
only to propose a positive {rational} number $\delta$ and a {rational}
lower-triangular matrix $L$. Exact arithmetic then forms
\begin{equation}
E=Q-\delta I-LL^{\mathsf T}
\label{eq:gram_residual}
\end{equation}
and bounds every absolute row sum using rational enclosures of $\sqrt2$.
Acceptance requires the strict exact inequality $\|E\|_\infty<\delta$. Since
$E$ is symmetric, for every $y\ne0$,
\begin{equation}
y^{\mathsf T}Qy
=\delta\|y\|_2^2+\|L^{\mathsf T}y\|_2^2+y^{\mathsf T}Ey
\ge(\delta-\|E\|_2)\|y\|_2^2
\ge(\delta-\|E\|_\infty)\|y\|_2^2>0,
\end{equation}
so the residual route is a proof of positive definiteness of the active
principal block~\cite{Rump06,Peyrl08}.

The residual test requires strict positivity, so it cannot certify singular
active blocks. If this test is inconclusive, the fallback performs an exact
fraction-free unpivoted $\mathrm{LDL}^{\mathsf T}$ elimination in $\mathbb F$.
A negative pivot is rejected. A zero pivot is accepted only when every
remaining entry in its Schur-complement column is exactly zero; under that
condition the zero direction decouples. Each positive pivot contributes a
nonnegative diagonal square. Induction over the columns therefore proves
$Q\succeq0$, including rank-deficient blocks.

{
The lower bound $C_*\ge293/100=2.930$ is certified using two different
exposure partitions (Table~\ref{tab:sos_comparison}).
The order-$6$ five-cell certificate uses the grid
\begin{equation*}
\left(0,\frac1{26},\frac13,\frac23,\frac{25}{26},1\right).
\end{equation*}
An independent order-$6$ six-cell certificate uses
\begin{equation*}
\left(0,\frac1{26},\frac14,\frac12,\frac34,\frac{25}{26},1\right).
\end{equation*}
}

The audit verifies all $2{,}665$ blocks of the order-$6$ five-cell
certificate ($332$ exact zero and $2{,}333$ residual-certified).
The certified objective $-n-V_1(0,\xv_0)=293/100$ yields
$C_*\ge293/100$ by the same integration argument as in
Theorem~\ref{thm:split_exact}.
Both archives, with audit logs and SHA-256 (Secure Hash Algorithm) checksum
manifests, are in the supplementary proof package
(Appendix~\ref{sec:methods}).

\subsection{\texorpdfstring{Order-7 certificate and exact verification}{Order-7 certificate and exact verification}}
\label{app:order7_certificate}
\leavevmode{
To tighten the order-$6$ bound, we construct an order-$7$ certificate
for~\eqref{eq:split_pop} using higher-degree potentials. It retains the quotient
algebra~\eqref{eq:quotient_algebra}, the terminal
map~\eqref{eq:terminal_functional}, and the three
cone constructions~\eqref{eq:occupation_cone}--\eqref{eq:terminal_cone};
only the order, the two end knots, and the target differ. The grid is
$(0,1/36,1/3,2/3,35/36,1)$, the ceiling is again $U=2935887/1000000$, the
potentials have degree at most $13$, and the archived target is
$L=5861/2000$. The memberships of~\eqref{eq:split_dual_conditions} are
therefore read with $\mathcal Q^{\mathrm{occ}}_7$,
$\mathcal Q^{\mathrm{int}}_7$, and $\mathcal H_7([0,1])$. Each cell
potential is stored as $776$ coefficients on the invariant normal
monomials. The serialized
Gram layout records $736$ occupation blocks per cell (largest side
$274$), $134$ interface blocks per internal knot (largest side $54$),
and two terminal blocks per cell (sides $8$ and $7$, corresponding to
$\mathbf a_7$ and the multiplier $\xi(1-\xi)\mathbf a_6$), $4{,}226$
blocks in all.

The exact audit certifies the lower bound $C_*\ge L=5861/2000=2.9305$.
The auditor of Appendix~\ref{app:auditor} regenerates this layout and
verifies $18{,}739$ coefficient identities over $\mathbb F$ with no
nonzero residual. Of the $4{,}226$ Grams, $506$ are exactly zero and
$3{,}720$ pass the residual route on their active principal blocks after removal of
identically zero rows and columns. Thus all Gram matrices are certified positive
semidefinite; $3{,}363$ of the original matrices are positive definite.
The exact $\mathrm{LDL}^{\mathsf T}$ fallback is never invoked, and the smallest
certified eigenvalue lower bound over the active principal blocks is
$246553280532227911527475802669/85070591730234615865843651857942052864$.
Both the
native audit and a separately executed eight-job independent audit
completed successfully.

}

\subsection{\texorpdfstring{Slack extension of the lower bound}{Slack extension of the lower bound}}
\label{app:slack_extension}
\leavevmode{
We strengthen the order-$7$ bound by using dynamical constraints to certify
a strictly positive lower bound on the total slack. The excitation and return
constraints exclude population pairs incompatible with the available exposure,
enabling sharper slack estimates. Let $C_T=C(S)=C$ denote the total exposure,
and consider a hypothetical pulse with $C_T\in[L,B]$, where
$L<B\le H=733/250<U$. We use the five-cell partition of $X(s)=C(s)/U$
defined in Appendix~\ref{app:order7_certificate}. Since $L>35U/36$, every such
trajectory crosses all four internal interfaces and terminates in the fifth cell,
with $X(S)\in[35/36,1)$. The certificate then gives the exact identity
\begin{equation}
C-L=S_{\mathrm{occ}}+S_{\mathrm{int}}+b_5\bigl(36C/U-35\bigr),
\label{eq:slack_identity}
\end{equation}
where $S_{\mathrm{occ}}$ and $S_{\mathrm{int}}$ are the cumulative non-negative occupation
and interface slacks:
\begin{equation}
S_{\mathrm{occ}} = \sum_{j=1}^5 \int_{s_{j-1}}^{s_j} q_j\bigl(\xi_j(s),\xv(s),u(s)\bigr)\,\dd s,
\qquad
S_{\mathrm{int}} = \sum_{j=1}^4 h_j\bigl(\xv(s_j)\bigr).
\label{eq:slack_sums_def}
\end{equation}
Here $0 = s_0 < s_1 < \dots < s_4 \le s_5 = S$ partition the active duration by the four
internal interface crossing times where $X(s_j) = a_j$ ($j=1,\dots,4$), $q_j(\xi,\xv,u) =
c(\xv) - L_j V_j(\xi,\xv) \ge 0$ is the cellwise drift slack, $h_j(\xv) = V_j(1,\xv) -
V_{j+1}(0,\xv) \ge 0$ is the interface jump polynomial, and $b_5 = \tau V_5 + n$ is the
fifth-cell terminal polynomial of~\eqref{eq:split_dual_conditions}. The local coordinate of
the terminating cell is $\xi_5(S) = 36C/U-35$. All three terms on the right-hand side
of~\eqref{eq:slack_identity} are non-negative; $b_5$ has degree at most $13$, its possible
degree-$14$ coefficient canceling identically.

The first barrier bounds the exposure required both to reach an intermediate
population pair and to return to zero excitation. Write
$p_k=|r_k|^2$ so that $c=p_1/2+p_2/4$. The sector equations give
$|p_1'|\le\sqrt{p_1(1-p_1)}$ and $|p_2'|\le\sqrt{2\,p_2(1-p_2)}$ almost
everywhere. With $a_0=1$, $a_{k+1}=a_k(2k+1)/(2k+2)$,
$S_n(p)=\sum_{k=0}^{n}a_kp^k$, and
$G_n(p)=p^{3/2}\sum_{k=0}^{n}[a_k/(2k+3)]p^k$, one has
$G_n'(p)=\sqrt{p}\,S_n(p)/2$, and exact polynomial arithmetic shows every
coefficient of $1-(1-p)S_n(p)^2$ nonnegative, so $S_n\sqrt{1-p}\le1$ on
$[0,1]$. The combination
\begin{equation}
W_n(p_1,p_2):=G_n(p_1)+\frac{G_n(p_2)}{2\sqrt2},
\qquad |W_n'|\le c,
\label{eq:w_barrier}
\end{equation}
vanishes at both CZ endpoints, hence
$W_n(s)\le\min\bigl(C(s),C_T-C(s)\bigr)$ throughout; the retained stages
use $n=8$ and $n=64$. A population rectangle is excluded exactly when the
lower-corner value of $W_n$ exceeds the radius
$R=\min(C_{\mathrm{right}},H-C_{\mathrm{left}})$ of its cost interval
$[C_{\mathrm{left}},C_{\mathrm{right}}]$,
decided by an exact rational comparison.

The second barrier uses the fact that both sectors are driven by the same field.
{We use the amplitude combination}
\begin{equation}
D=g_1r_2-\sqrt2\,g_2r_1,
\qquad D'=\frac{\ii}{2}\,u\,r_1r_2.
\label{eq:dark_coord}
\end{equation}
The exact identity
\begin{equation}
c^2-2|D'|^2=\frac{(2p_1-p_2)^2}{16}+\frac{(1-|u|^2)p_1p_2}{2}\ge0
\label{eq:dark_identity}
\end{equation}
gives $|D'|\le c/\sqrt2$. Since $D=0$ at both endpoints,
$|D(s)|\le\min\{C(s),C_T-C(s)\}/\sqrt2$.
Using $C_T\le H$, we obtain
$2|D(s)|^2\le\min\{C(s),H-C(s)\}^2$.
The reverse triangle inequality supplies the necessary population condition
\begin{equation}
|D|\;\ge\;\bigl|\sqrt{(1-p_1)p_2}-\sqrt{2(1-p_2)p_1}\bigr|.
\label{eq:dark_reverse}
\end{equation}
The expression inside the modulus decreases with $p_1$ and increases
with $p_2$, so on a closed box its relevant signed extremum sits at a
corner. Writing $\mu_1=(1-p_1)p_2$, $\mu_2=2(1-p_2)p_1$, and
$\nu=\mu_1+\mu_2-R^2/2$, a box is excluded exactly when the appropriate
strict sign of $\mu_1-\mu_2$ holds together with $\nu>0$ and
$\nu^2>4\mu_1\mu_2$; these are exact rational comparisons and equality
is retained.

We bound occupation slack on each fully traversed exposure interval
$I=[C_{\mathrm{left}},C_{\mathrm{right}}]$ within cell $j$, writing $\sigma_j=q_j$.
For each projection $m$ from an original Gram block $b(m)=b$, factoring the
common state-control monomial from selected basis coordinates leaves
$\boldsymbol v_m(\xi)=(c_i\xi^{k_i})_i$, with rational $c_i$ and nonnegative
integer $k_i$. {We embed a rational matrix $R_m$ on these coordinates so that
subtracting the embedded matrix from the original Gram block leaves a positive
semidefinite matrix. It defines}
$\beta_m(\xi)=\ell_m(\xi)\boldsymbol v_m(\xi)^\top R_m\boldsymbol v_m(\xi)${.}
{Here $\ell_m$ is the multiplier of the original Gram block in
Eq.~\eqref{eq:occupation_cone}: it is $1$ for the first family and $\xi(1-\xi)$
for the weighted families.} On each retained population
rectangle, nonnegative rational weights satisfy
$\sum_{m:b(m)=b}\lambda_m\le1$ for each original block, preventing double counting.

Physical accumulated exposure parameterizes the interval via the normalized coordinate
\begin{equation}
t = \frac{C(s) - C_{\mathrm{left}}}{C_{\mathrm{right}} - C_{\mathrm{left}}} \in [0, 1],
\qquad
\xi(t) = \frac{C_{\mathrm{left}} + (C_{\mathrm{right}} - C_{\mathrm{left}})t - U a_{j-1}}{U w_j}.
\label{eq:profile_coord_map}
\end{equation}
Along any trajectory, the allocated lower bound takes the form
\begin{equation}
\sigma_j\bigl(\xi(t),\xv,u\bigr) \ge N_I(t,p_1,p_2) := \sum_m \lambda_m \beta_m\bigl(\xi(t)\bigr)\,f_m(p_1,p_2),
\label{eq:NI_def}
\end{equation}
where $f_m(p_1,p_2) = (1-p_1)^{g_m} p_1^{r_m} (1-p_2)^{h_m} p_2^{s_m}$ are population
factors with nonnegative integer powers $g_m, r_m, h_m, s_m$ read from monomial
exponents. On every retained rectangle, we certify $N_I\ge d_Ic$ for a common profile
$d_I(t)=\sum_{k=0}^{14}\delta_{Ik}B_k^{14}(t)$,
$B_k^{14}(t)=\binom{14}{k}t^k(1-t)^{14-k}$,
with rational coefficients $\delta_{Ik}$. We check that $N_I-d_Ic$ has nonnegative
coefficients in the degree-$(14,7,7)$ tensor Bernstein basis in $t$ and the two
affinely normalized population coordinates. Since $\dd C=c\,\dd s$, integration
over a fully traversed exposure interval $I$ gives
\begin{equation}
\int_{\{s : C(s)\in I\}} \sigma_j\,\dd s \ge (C_{\mathrm{right}} - C_{\mathrm{left}})\sum_{k=0}^{14} \frac{\delta_{Ik}}{15}.
\end{equation}
{When strengthening an earlier bound, we count only the positive increase over
the profile integrals already included in that bound.}

Each mandatory internal interface crossing contributes additional certified slack.
At knot $j$, let $b$ index the original state-only Gram blocks $Q_{jb}$ of
the interface polynomial $h_j$. The certificate specifies nonnegative rational diagonal
matrices $M_{jb}$ on the corresponding monomial basis vectors $\boldsymbol{v}_{jb}(\xv)$
such that $Q_{jb} - M_{jb} \succeq 0$ is certified to be PSD. Summing diagonal
terms yields the pointwise lower bound
\begin{equation}
h_j(\xv) \ge N_j(p_1,p_2) := \sum_b \sum_i (M_{jb})_{ii} \bigl|(\boldsymbol{v}_{jb}(\xv))_i\bigr|^2,
\label{eq:Nj_def}
\end{equation}
whose squared monomial magnitudes are expressed as population factors via the sphere
normalizations; each original Gram block is counted once, and $N_j(p_1,p_2)$ is checked
against credited lower bounds on retained population rectangles.

The endpoint polynomial tests whether the credited slack together with the terminal
boundary term strictly exceeds $B-L$ for every candidate total exposure in $[L,B]$.
For a hypothetical pulse with total exposure $C_T \in
[L, B]$, we define the normalized endpoint variable
\begin{equation}
\eta = \frac{C_T - L}{B - L} \in [0, 1], \qquad C_T = L + (B - L)\eta.
\label{eq:endpoint_eta_def}
\end{equation}
With the scale ratio $\rho = (B - L)/(B_0 - L)$, the predecessor endpoint coordinate is
$\eta_0 = (C_T - L)/(B_0 - L) = \rho\eta$. If the predecessor stage certified lower bound
$B_0$ via an endpoint polynomial $P_{\mathrm{old}}$ of degree at most $13$, its base form
is
\begin{equation}
P_{\mathrm{old}}(\eta_0) = K_{\mathrm{old}} + b_5\left(\frac{36[L + (B_0 - L)\eta_0]}{U} - 35\right) - (B_0 - L),
\label{eq:Pold_base_def}
\end{equation}
where $K_{\mathrm{old}}$ is the sum of all previously credited non-negative occupation and
interface slacks. The updated degree-at-most-$13$ endpoint polynomial is
\begin{equation}
P_{\mathrm{new}}(\eta) = P_{\mathrm{old}}(\rho\eta) + \Delta - (B - B_0),
\label{eq:endpoint_compose}
\end{equation}
where $\Delta$ is the increment of newly credited slack over $K_{\mathrm{old}}$.
Identity~\eqref{eq:endpoint_compose} is verified identically at $14$ distinct rational
nodes, and strict positivity of $P_{\mathrm{new}}$ on $[0,1]$ is read from its $14$
Bernstein coefficients in exact arithmetic. A hypothetical pulse with $C_T \le B$ would
force the right-hand side of~\eqref{eq:slack_identity} to strictly exceed $B - L$, which
contradicts the identity; hence $C_* \ge B$ holds without an attainment assumption or
duration cap.

The checkers evaluate every
coefficient inequality and Bernstein coefficient in $\mathbb Q$, while
the underlying SDP identities remain in $\mathbb Q(\sqrt2)$ as pairs of
rationals; {rigorous interval enclosures computed with Arb~\cite{Johansson17}}
are used in the interface
$\mathrm{LDL}^{\mathsf T}$ semidefinite test and the matrix-order tests
for occupation projections.
Verification methods are summarized in Appendix~\ref{sec:methods};
detailed counts and test results are reported in the archived logs. Each stage's checker pins its predecessor certificates by
SHA-256, and the reproduction scripts replay their verifiers. The packaged validation
also re-executes the inherited global source audit, reproducing its retained output
byte for byte~\cite{repo_pub}.
}

\section{Nonlocal phase velocity and synchronization error}
\label{app:coordinates}

The nonlocal phase is the two-atom ground-state phase minus twice the
single-atom ground-state phase.
We derive how this phase difference changes under the shared laser drive.
Because the collective coupling is enhanced by $\sqrt{2}$, we compare the
amplitude ratio $r_2/g_2$ with $\sqrt{2}$ times $r_1/g_1$.

\begin{lemma}[Nonlocal phase velocity and synchronization error]
\label{lem:ratio}
In each sector $k \in \{1, 2\}$, define stereographic coordinates
$z_k := r_k/g_k$ for $g_1 g_2 \ne 0$, distinct from the overlap
$z(\xv)$ of Appendix~\ref{sec:analytical}.
Let $w := z_2 - \sqrt{2} z_1$ denote the synchronization error, which measures
departure from the scaled common response, and let $\zeta := \Phi_2 - 2\Phi_1$
be the nonlocal phase ($\arg g_2 - 2\arg g_1$).
With overdots denoting differentiation with respect to active time $s$, the
velocities satisfy
\begin{equation}
\dot{\zeta}
\;=\;
-\frac{1}{\sqrt{2}}\operatorname{Re}(u w),
\qquad
\dot{w}
\;=\;
\frac{\ii}{\sqrt{2}}\,u(s)\bigl(z_2^2-z_1^2\bigr).
\label{eq:ratio}
\end{equation}
\end{lemma}
\begin{proof}
Whenever $g_1g_2\ne0$, Eq.~\eqref{eq:schrod} gives the Riccati equations
\begin{equation}
\dot z_k \;=\; -\frac{\ii\sqrt{k}}{2}u^*
+\frac{\ii\sqrt{k}}{2}u z_k^2.
\end{equation}
The terms independent of $z_k$ cancel in $\dot z_2-\sqrt{2}\dot z_1$, yielding the second
identity in~\eqref{eq:ratio}. Moreover
\begin{equation}
\dot\Phi_k
\;=\;
-\frac{\sqrt{k}}{2}\operatorname{Re}(u z_k),
\end{equation}
so
\begin{equation}
\dot\zeta
\;=\;
\operatorname{Re}(u z_1)
\;-\;
\frac{1}{\sqrt{2}}\operatorname{Re}(u z_2)
\;=\;
-\frac{1}{\sqrt{2}}\operatorname{Re}(u w),
\end{equation}
which is the first identity in~\eqref{eq:ratio}.
\end{proof}

\section{Asymptotic completeness of the real-coefficient moment-SOS hierarchy}
\label{app:sos-convergence}

We establish asymptotic completeness of the exposure-partitioned moment-SOS hierarchy
for any fixed partition. {Let $L_d^{\mathrm{unsplit}}$ and $L_d^{\mathrm{split}}$
denote the optimal values of the stationary and exposure-partitioned order-$d$
programs, respectively.} Feasible stationary certificates embed as the
cell-independent special case
\[
W_j(\xi,\xv)=W(\xv)\qquad\text{for every cell }j,
\]
which yields
\[
L_d^{\mathrm{unsplit}}\le L_d^{\mathrm{split}}\le C_*.
\]
It therefore suffices to prove that $L_d^{\mathrm{unsplit}}\to C_*$ as
$d\to\infty$. Throughout, both hierarchies use complete polynomial spaces
with real coefficients.

\subsection{Stationary hierarchy}

We adopt the dynamics $\dot{\xv}=f_u(\xv)$, running cost
$c$, initial state $\xv_0$, and target set
$Z:=\mathcal X_{\mathrm{CZ}}$ from Sec.~\ref{sec:moments_sos}. Let
$\mathcal M:=\mathcal X$ denote the state manifold and
$F=\{\xv\in\mathcal M:r_1=r_2=0\}$ its zero-cost {set}. We
write $f_uW=\nabla W\cdot f_u$ for the derivative of $W$ along $f_u$.
State and control normalization constraints generate the ideal
\begin{equation}
\begin{aligned}
\mathcal I=\langle &|g_1|^2+|r_1|^2-1,\\
&|g_2|^2+|r_2|^2-1,\ |u|^2-1\rangle,
\end{aligned}
\end{equation}
and $G$ denotes the symmetry group expressed in real
coordinates.

The stationary order-$d$ program maximizes $W(\xv_0)$ over real,
$G$-invariant polynomials with $\deg W\le2d-1$, subject to
\begin{equation}
\begin{gathered}
 W|_Z\le0,\qquad \deg h_\ell\le d,\\
 c+f_uW\equiv\sum_\ell h_\ell^2\pmod{\mathcal I}.
\end{gathered}
 \label{eq:erev-stationary}
\end{equation}
We denote its optimal value by $L_d^{\mathrm{unsplit}}$.
Integrating $c+f_uW\ge0$ along any driven trajectory and applying
$W|_Z\le0$ shows that $W(\xv_0)$ bounds the exposure of every successful
pulse from below, so $L_d^{\mathrm{unsplit}}\le C_*$.

\begin{theorem}[Stationary completeness]\label{thm:asymptotic_completeness}
For the globally driven, perfect-blockade CZ problem with measurable phase
controls, exact terminal closure, and arbitrary finite active duration,
\begin{equation}
\begin{gathered}
 0\le L_d^{\mathrm{unsplit}}\le L_{d+1}^{\mathrm{unsplit}}\le C_*,\\
 \lim_{d\to\infty}L_d^{\mathrm{unsplit}}=C_*.
\end{gathered}
 \label{eq:erev-convergence}
\end{equation}
\end{theorem}

The proof proceeds by constructing nearly optimal smooth
bounding functions and then approximating them with polynomial
SOS certificates.

\subsection{Existence of nearly optimal smooth bounds}

We seek a smooth state-dependent function $W_0$ that vanishes at the CZ target and decreases
by no more than the accumulated Rydberg exposure along any pulse. Its initial value then
bounds the required exposure from below. To establish completeness, we must obtain
$|f_uW_0|\le c$, $W_0|_Z=0$, and $W_0(\xv_0)>a$ for every $a<C_*$.
The physical obstacle in constructing $W_0$ via occupation measures in active time $s$
is that small Rydberg populations permit trajectories of arbitrarily long duration $S$
at bounded exposure.
Because the total mass of an occupation measure in active time equals $S$, a bounded exposure
budget imposes no uniform mass bound as $S\to\infty$, so the usual compactness argument for finite measures does not apply.

We bypass this by parameterizing evolution by accumulated exposure rather than time.
The resulting velocity $f_u/c$ is singular where the Rydberg population vanishes.
We choose new coordinates in which the velocity per unit exposure stays bounded and extends
continuously to this zero-population set.
We optimize over exposure records that describe a flow through state space along the
allowed driven motion. The flow enters at the initial state and exits at the CZ target.
Away from these endpoints, incoming flow balances outgoing flow: nothing appears or
disappears along the way.

For example, half the flow can follow one successful pulse's route and half can follow
another's. We allow these split flows without requiring a single pulse to reproduce the
entire record. We then prove that, for every allowed record and every $\varepsilon>0$,
there exists a finite-duration pulse that reaches the CZ target exactly. Its total
exposure is at most the record's total exposure plus $\varepsilon$. Allowing general
records therefore cannot lower the minimum exposure. Convex separation then yields the
smooth bound $W_0$.

\subsubsection*{1. Exposure coordinates}

We define the exposure clock
\[
t(s)=\int_0^s c(\xv(s'))\,\dd s',
\]
so that $t(S)=C$.
Where $c(\xv)>0$, the transformed velocity field is $\dd\xv/\dd t=f_u(\xv)/c(\xv)$.
This expression becomes singular on the zero-cost set $F=\{\xv\in\mathcal M:r_1=r_2=0\}$.
To remove this singularity, we introduce coordinates $J$ whose velocity field extends continuously
to $J(F)$ and remains bounded.

We permit controls in the unit disk $U=\{u\in\mathbb C:|u|\le1\}$. By Lemma~\ref{lem:bang},
each weaker-drive pulse can be traversed at full amplitude without increasing exposure, so admitting
weaker amplitudes cannot reduce $C_*$.
Recall the observable $z$ of Appendix~\ref{sec:analytical}:
\begin{equation}
\begin{gathered}
 z=\bar g_2g_1^2+\sqrt2\,\bar r_2g_1r_1,\\
 f_uz=-\frac{\ii}{\sqrt2}\bar r_2ur_1^2,\\
 |f_uz|\le\kappa c^{3/2},
\end{gathered}
 \label{eq:erev-progress}
\end{equation}
for a fixed $\kappa>0$. With $x_{\mathbb R}\in\mathbb R^8$ denoting the real state
coordinates, define
\begin{equation}
\begin{gathered}
 J(\xv)=(\Re z,\Im z,c^2,c^2x_{\mathbb R}),\qquad Y=J(\mathcal M),\\
 B(J(\xv),u)=\frac{DJ(\xv)f_u(\xv)}{c(\xv)}\quad(c>0),
\end{gathered}
 \label{eq:erev-compactification}
\end{equation}
{where $DJ$ is the Jacobian matrix of $J$. We set $B=0$ on $J(F)$.}
The state space $Y$ is compact because $\mathcal M$ is compact
and $J$ is continuous. The map $J$ is injective off $F$, collapses $F$ to a circle, and maps the initial
state $\xv_0$ and the target set $Z$ to $y_0=(1,0,\boldsymbol 0)$ and $y_\pi=(-1,0,\boldsymbol 0)$,
respectively. Since $|f_uc|\le\sqrt c$ and \eqref{eq:erev-progress} holds, $|B|=O(\sqrt c)$. Hence $B$,
representing velocity per unit exposure, is a continuous, bounded vector field on $Y\times U$, real-linear in $u$.
These properties enable physical trajectory recovery in Step~3,
control averaging in Step~4, and convex separation in Step~5.

\subsubsection*{2. Exposure measures}

For any successful physical pulse $\gamma$ transferring $\xv_0$ to $Z$ in active time $S$,
each interval contributes $c(\xv(s))\,\dd s$ at the instantaneous state and control.
Its exposure record is the nonnegative measure $\nu_\gamma$ on $Y\times U$ defined by
\[
\begin{gathered}
\nu_\gamma=\int_0^S c(\xv(s))
\delta_{(J(\xv(s)),u(s))}\,\dd s,\\
\nu_\gamma(Y\times U)=C,
\end{gathered}
\]
where $\delta_{(y,u)}$ denotes the point mass at $(y,u)$, and total mass equals total exposure $C$.
For every smooth observable $\psi\in C_c^\infty(\mathbb R^{11})$ with compact support, the integrated
chain rule along the trajectory yields the net balance identity
\[
\begin{aligned}
\int_{Y\times U} B\cdot\nabla\psi\,\dd\nu_\gamma
&=\int_0^S \frac{\dd}{\dd s}\psi(J(\xv(s)))\,\dd s\\
&=\psi(y_\pi)-\psi(y_0).
\end{aligned}
\]
Relaxing the requirement that $\nu$ be generated by a physical pulse motivates the optimization:
\begin{equation}
\begin{gathered}
 p=\inf_{\nu\ge0}\nu(Y\times U)\quad\text{subject to}\\
 \int B\cdot\nabla\psi\,\dd\nu=\psi(y_\pi)-\psi(y_0),\\
 \forall\psi\in C_c^\infty(\mathbb R^{11}).
\end{gathered}
 \label{eq:erev-measures}
\end{equation}
In \eqref{eq:erev-measures}, $\nu$ is an arbitrary finite nonnegative measure on $Y\times U$,
with no requirement that it arise from a single physical pulse.
Because every physical pulse induces a feasible measure with mass equal to its exposure, the relaxed infimum
satisfies $p\le C_*$.

Proving the reverse inequality $p\ge C_*$ (and hence $p=C_*$) requires showing that every feasible exposure measure supplies a physical pulse of finite active duration whose exposure cost is at most the measure's total mass plus an arbitrarily small $\varepsilon$.
This bridge does not require reproducing the entire measure with a single pulse: Step~4 extracts a single finite-exposure trajectory from the measure, and Step~3 converts that trajectory into a finite-active-duration physical pulse reaching the target set exactly with exposure cost at most the trajectory duration plus $\varepsilon$.

\subsubsection*{3. Finite-duration realization}

While Step~2 admits generalized exposure records (measures) that describe split or combined flows through state space, and Step~4 extracts an individual exposure-parametrized trajectory from such a measure, Step~3 supplies the auxiliary reconstruction from this trajectory to a true physical pulse.
Given an absolutely continuous trajectory $y(t)$ solving $\dot y=B(y,u)$ in $Y$ from $y_0$ to $y_\pi$ under an admissible measurable control $u(t)\in U$ with finite exposure-clock parameter length $T$, Step~3 constructs for each $\varepsilon>0$ a physical pulse of finite active duration $S$ that starts at $\xv_0$, reaches the target set $Z$ exactly, and incurs exposure cost at most $T+\varepsilon$.

The physical obstacle is the disparity between active duration $S$ (elapsed laser interaction time) and the exposure-clock duration $T$ (the available exposure budget).
Along a physical trajectory, the physical clock $s$ and exposure clock $t$ are related by $\dd t = c\,\dd s$, or $\dd s = \dd t/c$ where the running cost $c$ is positive.
Because the running cost vanishes on the unexcited set $F$, the physical rate $\dd s/\dd t = 1/c$ diverges near zero cost.
An exposure-parametrized trajectory spending finite exposure $T$ near $c=0$ could therefore accumulate infinite active duration if naively lifted to physical time.

To guarantee finite active duration at exposure cost near $T$, we reconstruct the pulse in three stages:
(i) retain sizable excursions where $c$ is large enough that physical time accumulates at a bounded rate,
(ii) replace near-zero-cost stretches by short local steering maneuvers between projected endpoints on the zero-cost submanifold $F$, and
(iii) repair the small conditional phase omitted during the excised segments with a final steering correction.

These local steering maneuvers rely on the controllability of the physical system $\dot\xv = f_u(\xv)$.
The complex drive $u = u_x + \ii u_y$ provides two independent control quadratures corresponding to the two laser field phases.
Although these controls directly generate only two velocity directions on the six-dimensional state manifold $\mathcal M\cong SU(2)\times SU(2)$, alternating their flows generates motion along the remaining directions via Lie brackets (commutators).
Bracket layers of lengths 1 through 4 have cumulative span dimensions $(2,3,5,6)$, fully spanning the six-dimensional tangent space:
sector weights at lengths 1 and 3 separate the two control quadratures, while lengths 2 and 4 separate the diagonal generators.
The uniform ball-box theorem on compact manifolds~\cite{Jean14} establishes a quantitative local steering bound:
nearby states at Riemannian distance $d_R$ can be connected by an admissible control in active time $S=O(d_R^{1/4})$.

Near the zero-cost submanifold $F=\{\xv\in\mathcal M:r_1=r_2=0\}$, steering requires very little exposure.
Starting on $F$, the sector bounds $|r_k(s)|\le\sqrt{k}s/2$ for $k\in\{1,2\}$ constrain the running cost to $c(s)\le s^2/4$.
Integrating this running cost over active duration $S$ bounds the accumulated exposure by $\int_0^S(s^2/4)\,\dd s \le S^3/12$.
Combining this cubic scaling with the ball-box active time $S=O(d_R^{1/4})$, steering from a point on $F$ to a nearby state at distance $d_R$ incurs an exposure cost of at most $O(d_R^{3/4})$.
Because the physical vector field satisfies $f_{-u}=-f_u$, time reversal under control $-u$ steers from a state at distance $d_R$ into $F$ at the identical exposure cost $O(d_R^{3/4})$.

We isolate sizable excursions using two cutoffs $0<a_0<b/2$: $b$ decides which excursions are kept, and $a_0$ sets their truncation endpoints away from $c=0$.
Along positive-cost segments, $|\dd(c^{3/2})/\dd t|\le3/2$, extending continuously across zeros.
An excursion is a connected time interval where $c(t)>a_0$; we retain those whose peak running cost exceeds $b$.
Because every retained excursion rises from $a_0$ past $b$ and falls back, it consumes an exposure duration of at least a constant times $b^{3/2}$, bounding the number of retained excursions by $N=O(Tb^{-3/2})$.
Off $F$, dividing the final eight coordinates of $J(\xv)$ by its third coordinate $c^2$ recovers the real coordinates $x_{\mathbb R}$, making $J^{-1}$ smooth on $\{c\ge a_0\}$.
Each retained segment therefore lifts via $\xv(t)=J^{-1}(y(t))$ and $\dd s=\dd t/c$ to physical dynamics $\dd\xv/\dd s=f_u(\xv)$, with total retained active duration at most $T/a_0$.

At each retained endpoint, the running cost satisfies $c=a_0$, so excitation amplitudes are of order $O(\sqrt{a_0})$.
We project each endpoint onto $F$ using
\[
 P(\xv)=\left(\frac{g_1}{|g_1|},0,
 \left(\frac{g_1}{|g_1|}\right)^2\frac{\bar z}{|z|},0\right)\in F
\]
which preserves the phase angle $\arg z$ while resetting Rydberg excitations to zero.
The Riemannian distance between $\xv$ and $P(\xv)$ is $d_R=O(\sqrt{a_0})$.
Steering from $P(\xv)$ into the excursion's initial state, and from its final state into $P(\xv)$, incurs an exposure cost of $O(d_R^{3/4})=O(a_0^{3/8})$ per retained excursion.
These endpoint maneuvers convert each retained segment into a physical block that begins and ends on $F$ in finite active time.

Across the discarded intervals, the running cost remains bounded by $c\le b$.
For sufficiently small $b$, $|z|\ge1/2$, and the Step~1 observable bound $|\dd z/\dd t|\le\kappa c^{1/2}$ yields $|\dd\arg z/\dd t|\le2\kappa\sqrt b$.
The total conditional phase omitted across all discarded intervals is bounded by $|E|=O(T\sqrt b)$, where $E$ is the signed net phase deficit.
On $F$, the observable reduces to $z=\bar g_2 g_1^2$, so $\arg z = 2\arg g_1 - \arg g_2$ tracks the conditional gate phase with opposite sign.
By symmetry, multiplying the amplitudes $(g_k,r_k)$ in sector $k$ by a constant phase $\mathrm{e}^{\ii\phi_k}$ maps any solution of $\dot\xv=f_u(\xv)$ to another valid solution driven by the same control with identical running cost.
Choosing these constant phase offsets block-by-block matches the entry state of each $F$-to-$F$ block to the exit state of its predecessor (and the initial block to $\xv_0$), exactly preserving each block's internal conditional-phase increment.
Because the chained trajectory terminates on $F$ and the target set $Z$ consists of states on $F$ with $\arg z=\pi\pmod{2\pi}$, the remaining conditional phase deficit corresponds to Riemannian distance $d_R=O(|E|)=O(T\sqrt b)$ on $F$.
A final steering correction from $F$ to $Z$ closes this gap exactly at exposure cost $O(|E|^{3/4})=O(T^{3/4}b^{3/8})$.
Summing the retained exposure, the $2N$ endpoint connections, and the terminal phase repair yields
\begin{equation}
 C_{a_0,b}\le T+O(Tb^{-3/2}a_0^{3/8})
                  +O(T^{3/4}b^{3/8}).
 \label{eq:erev-patching}
\end{equation}
Setting $a_0=b^5$ makes both additional exposure terms vanish as $b\to0$ for any fixed $T$.
Given $\varepsilon>0$, choosing $b$ sufficiently small ensures $C_{a_0,b}\le T+\varepsilon$.
Because $a_0>0$ and $b>0$ are strictly positive, the pulse consists of finitely many retained blocks ($N<\infty$) and finitely many finite-time steering connections, guaranteeing that the pulse has finite active duration $S$ (no uniform bound on $S$ as $\varepsilon\to0$ is required).
Finally, Lemma~\ref{lem:bang} converts intermediate controls $|u|<1$ to unimodular controls $|u|=1$ without increasing exposure, completing the finite-active-duration physical pulse for Step~4.

\subsubsection*{4. Equality of the two infima}

We now show that from any feasible measure $\nu$ in \eqref{eq:erev-measures}, one can extract
a continuous trajectory in $Y$ whose exposure cost does not exceed the mass $\nu(Y\times U)$.
We disintegrate $\nu$ as $\nu(\dd y,\dd u)=\rho(\dd y)\nu_y(\dd u)$, where the spatial marginal $\rho$
records exposure spent near state $y$, and the conditional probability measure $\nu_y$ describes the
distribution of controls applied at $y$. Defining the conditional average control $\bar u(y)=\int_U u\,\dd\nu_y(u)$,
convexity of $U$ ensures $\bar u(y)\in U$. By real-linearity of $B(y,u)$ in $u$, the averaged velocity is
\[
\begin{aligned}
v(y)&=\int_U B(y,u)\,\dd\nu_y(u)\\
&=B(y,\bar u(y)),
\end{aligned}
\]
which is generated by an admissible control in the disk. We choose a {Borel-measurable representative} of $\bar u$ on $Y$,
extending it by zero on the $\rho$-null exceptional set.

The vector measure $v\rho$ defines a flow with a unit source at $y_0$ and a unit sink at $y_\pi$.
{Formally, $v\rho$ is a normal 1-current: both its mass $\int_Y |v|\,\dd\rho$
and the mass of its boundary $\delta_{y_\pi}-\delta_{y_0}$ are finite.}
Proposition~3.8 of Paolini and Stepanov~\cite{Paolini13} removes cycles without altering the boundary,
yielding an acyclic subcurrent $\lambda v\rho$ with $0\le\lambda(y)\le1$.
{Here acyclic means that the retained current contains no nonzero subcurrent with zero boundary.}
Applying their arc-decomposition
theorem (Theorem~5.1) and localized mass identity (Lemma~4.17) decomposes this acyclic flow into an ensemble
of directed arcs $\gamma$ connecting $y_0$ to $y_\pi$, tangent to $v/|v|$ almost everywhere, distributed
according to a probability measure $\eta$ satisfying
\begin{equation}
\begin{gathered}
 \int T_\gamma\,\dd\eta(\gamma)
 =\int_{\{|v|>0\}}\lambda\,\dd\rho\le\nu(Y\times U),\\
 T_\gamma=\int_\gamma\frac{\dd\ell}{|v|}.
\end{gathered}
 \label{eq:erev-weightedmass}
\end{equation}
Here $T_\gamma$ represents the exposure duration along the arc $\gamma$ after
reparameterization, with the inverse speed $|v|^{-1}$ converting arc length $\dd\ell$ to exposure duration;
\eqref{eq:erev-weightedmass} follows from truncating this weight in the localized mass identity.
Since the expectation of $T_\gamma$ under $\eta$ is at most $\nu(Y\times U)$, there exists an arc $\gamma$
with $T_\gamma\le\nu(Y\times U)$. Its exposure clock $\tau(\ell)=\int_0^\ell |v(\gamma(q))|^{-1}\dd q$ is
absolutely continuous with Lipschitz inverse because $B$ is bounded, so the reparameterized arc solves
$\dot y=B(y,\bar u(y))$ almost everywhere.

Applying the finite-duration realization of Step~3 to this trajectory produces, for each
$\varepsilon>0$, a finite-active-duration pulse with exact terminal closure on $Z$ satisfying
\[
\begin{aligned}
C_*&\le C_\varepsilon\le T_\gamma+\varepsilon\\
&\le\nu(Y\times U)+\varepsilon.
\end{aligned}
\]
Taking $\varepsilon\to0$ and the infimum over all feasible measures $\nu$ yields $C_*\le p$. Combined with
$p\le C_*$ from Step~2, this establishes $p=C_*$.

\subsubsection*{5. Smooth lower bounds}

Because $p=C_*$, any threshold $a<C_*$ cannot be achieved by any feasible
measure in \eqref{eq:erev-measures}. We exploit this infeasibility by separating the target
endpoint balance and budget from the cone of attainable changes and budgets:

\begin{lemma}[Smooth subsolutions]\label{lem:erev-smooth}
For every $a<C_*$ there is $\varphi\in C_c^\infty(\mathbb R^{11})$ such that
\begin{equation}
\begin{gathered}
 W_0=\varphi\circ J-\varphi(y_\pi),\\
 |f_uW_0|\le c,\qquad W_0|_Z=0,\qquad W_0(\xv_0)>a.
\end{gathered}
 \label{eq:erev-smooth}
\end{equation}
\end{lemma}

\begin{proof}
{Let $\mathcal D=C_c^\infty(\mathbb R^{11})$ be the space of smooth, compactly supported
test functions, and let $\mathcal D'$ denote the space of continuous linear functionals on $\mathcal D$.
We equip $\mathcal D'$ with the weak topology $\sigma(\mathcal D',\mathcal D)$,
in which convergence is tested on every $\psi\in\mathcal D$.}
{Define} $A\nu(\psi)=\int B\cdot\nabla\psi\,\dd\nu$.
Consider the convex cone of attainable net changes paired with allowed exposure budgets:
\[
 \mathcal C=\{(A\nu,m):\nu\ge0,\ m\ge\nu(Y\times U)\}
 \subset \mathcal D'\times\mathbb R.
\]
The cone $\mathcal C$ is closed in the weak topology $\sigma(\mathcal D',\mathcal D)\times\mathbb R$:
a bound on the budget $m$ bounds the total mass of $\nu$, bounded positive measures on compact $Y\times U$
have weak-* convergent subnets, and continuity of $B$ preserves the integrals defining $A$.
The target pair $(\delta_{y_\pi}-\delta_{y_0},a)$ lies strictly outside $\mathcal C$ because any feasible
record requires mass at least $p=C_*>a$. By the Hahn--Banach separation theorem, there exists $\lambda_0\ge0$
and $\varphi\in\mathcal D$ strictly separating this point from $\mathcal C$:
\[
\lambda_0m+A\nu(\varphi)\ge0\quad\forall(A\nu,m)\in\mathcal C,
\]
while $\lambda_0a+\varphi(y_\pi)-\varphi(y_0)<0$. Feasible physical measures exclude $\lambda_0=0$, so
scaling allows setting $\lambda_0=1$.

Evaluating on atomic measures $\nu=\delta_{(y,u)}$ with budget $m=1$ gives $1+B(y,u)\cdot\nabla\varphi(y)\ge0$.
Control reversal $u\mapsto -u$ negates $B$, yielding $|B\cdot\nabla\varphi|\le1$. Moreover, strict separation
gives $\varphi(y_0)-\varphi(y_\pi)>a$.

Setting $W_0=\varphi\circ J-\varphi(y_\pi)$, we have $W_0|_Z=\varphi(y_\pi)-\varphi(y_\pi)=0$
and $W_0(\xv_0)=\varphi(y_0)-\varphi(y_\pi)>a$. The chain rule yields
\[
\begin{aligned}
f_uW_0(\xv)
&=\nabla\varphi(J(\xv))\cdot DJ(\xv)f_u(\xv)\\
&=c(\xv)\,B(J(\xv),u)\cdot\nabla\varphi(J(\xv)).
\end{aligned}
\]
Since $|B\cdot\nabla\varphi|\le1$, multiplying by $c(\xv)\ge0$ gives $|f_uW_0|\le c$, proving
\eqref{eq:erev-smooth}.
\end{proof}

\subsection{Polynomial SOS certificates}

The previous subsection established smooth bounds arbitrarily close to $C_*${,
 the infimum of the total exposure over successful CZ pulses}.
{
Recall that $\xv=(g_1,r_1,g_2,r_2)$ collects the ground and Rydberg amplitudes
of the two driven sectors, and $\xv_0=(1,0,1,0)$ is the initial state.
The exposure rate is $c(\xv)=(2|r_1|^2+|r_2|^2)/4$.
For a candidate bound $W(\xv)$, the notation $f_uW=\nabla W\cdot f_u$
denotes its derivative along the physical dynamics
$\dd\xv/\dd s=f_u(\xv)$, with $s$ the active time and $u$ the common drive control.
}
To obtain a feasible point of the stationary SOS program, we must first
replace a smooth bound by a polynomial $W$ without violating
$c+f_uW\ge0$, and then express this nonnegative residual as a sum of
polynomial squares on the normalized state--control set. The first step
below constructs $W$. The remaining two steps certify its residual
without changing $W$ or its initial value.

{
The group $G$ comprises the phase rotations and reflection introduced in
Sec.~\ref{sec:sos_reductions}; a $G$-invariant polynomial assigns the same
value to symmetry-related states.
}

\begin{lemma}[Polynomial certificates]\label{lem:erev-polynomial}
For every $\varepsilon>0$, a real, $G$-invariant stationary polynomial is
feasible in \eqref{eq:erev-stationary} at some finite order and satisfies
$W(\xv_0)>C_*-\varepsilon$.
\end{lemma}

\begin{proof}
\emph{1.~Approximate the smooth bound while preserving its rate inequality.}
{
The polynomial map $J$ of \eqref{eq:erev-compactification} sends the physical
state $\xv$ to transformed coordinates $y=J(\xv)\in\mathbb R^{11}$;
$Y$ denotes the set of these transformed physical states.
The function $\varphi(y)$ is the smooth real-valued function supplied by
Lemma~\ref{lem:erev-smooth}. The composition $\varphi\circ J$ means evaluating
$\varphi$ at $J(\xv)$. All states in the CZ target set $Z$ map to the same
point $y_\pi$, so subtracting $\varphi(y_\pi)$ sets the target value to zero.
}
Choose the smooth bound
$W_0=\varphi\circ J-\varphi(y_\pi)$ of
Lemma~\ref{lem:erev-smooth} sufficiently close to $C_*$.
Approximating its values alone is insufficient: if
$\widetilde W=W_0+e$, the derivative error $f_ue=\nabla e\cdot f_u$
changes the residual according to
\[
c+f_u\widetilde W=(c+f_uW_0)+f_ue.
\]
Even uniformly small errors in both values and first derivatives can
violate nonnegativity near a state where the exposure rate $c$ vanishes.
For example, consider auxiliary scalar dynamics $\dd x/\dd s=1$ on
$[-1,1]$, with exposure rate $c_{\rm toy}(x)=x^2$ and bound
$W_{\rm toy}(x)=0$.
The approximation $\widetilde W_{\rm toy}(x)=-\delta x$, $\delta>0$,
has value and derivative errors at most $\delta${. Along the
trajectory $x=x(s)$, its derivative is
$\dd\widetilde W_{\rm toy}(x(s))/\dd s
=-\delta\,\dd x/\dd s=-\delta$, so}
\[
c_{\rm toy}+\frac{\dd\widetilde W_{\rm toy}}{\dd s}
=x^2-\delta<0
\qquad (|x|<\sqrt{\delta}).
\]
Reducing $\delta$ narrows the region of failure without eliminating it.
For the physical problem, the exposure rate $c$ likewise vanishes on
$F=\{r_1=r_2=0\}$, so a fixed absolute bound on the derivative error
does not suffice.

{
We instead approximate $\varphi$ in the transformed coordinates $y=J(\xv)$.
Here $DJ$ is the Jacobian matrix of $J$, and $B(y,u)$ is the transformed
velocity per unit of accumulated exposure. Along active time $s$, the
chain rule gives
\[
\frac{\dd}{\dd s}J(\xv(s))
=DJ(\xv)f_u(\xv)=c(\xv)B(J(\xv),u).
\]
The preceding subsection established that $B$ is bounded even where
$c$ approaches zero. Thus differentiating an approximation error in
these coordinates introduces a factor of $c$, as quantified below.
}
Choose polynomials $P_m$ on a compact box containing $Y$ with
$\|P_m-\varphi\|_{C^1}\le\eta_m\to0$, where the $C^1$ norm controls
values and first derivatives{. The smooth-bound construction
uses the control disk $U=\{u\in\mathbb C:|u|\le1\}$.
Set $K_B=\sup_{Y\times U}|B|$, the finite uniform bound on the transformed
velocity over these states and controls}.
The chain rule gives
\[
\begin{gathered}
f_u(P_m\circ J-W_0)=c\,B\cdot\nabla(P_m-\varphi),\\
|f_u(P_m\circ J-W_0)|\le cK_B\eta_m,\\
|f_u(P_m\circ J)|\le c(1+K_B\eta_m).
\end{gathered}
\]
Here $B$ is evaluated at $(J(\xv),u)$ and the gradients at $J(\xv)$.
The error now vanishes wherever $c=0$. Subtracting the target value
and dividing by $1+K_B\eta_m$ therefore yield the polynomial
\begin{equation}
\begin{gathered}
 Q_m=\frac{P_m\circ J-P_m(y_\pi)}{1+K_B\eta_m},\\
 |f_uQ_m|\le c,\qquad Q_m|_Z=0.
\end{gathered}
 \label{eq:erev-approximation}
\end{equation}
Since $J(Z)=\{y_\pi\}$, the subtraction enforces $Q_m|_Z=0$;
the division restores the exact derivative bound. As $m\to\infty$,
$Q_m(\xv_0)$ approaches $W_0(\xv_0)$.

To leave a positive margin away from $F$, multiply $Q_m$ by
$1-\alpha$, with $0<\alpha<1$. This scales its derivative by the
same factor:
\[
\begin{aligned}
c+f_u[(1-\alpha)Q_m]
&=c+(1-\alpha)f_uQ_m\\
&\ge c-(1-\alpha)c=\alpha c.
\end{aligned}
\]
The initial value is also multiplied by $1-\alpha$, so the loss in
the lower bound can be made arbitrarily small by taking $m$ large
and $\alpha$ small. Average $(1-\alpha)Q_m$ over $G$ and call the
result $W$. {Every state in the symmetry orbit $G\xv_0$
maps to the same transformed initial point $y_0=J(\xv_0)$, so
$J(G\xv_0)=\{y_0\}$ and averaging preserves the initial value.}
Equivariance of the dynamics and invariance of
$c$ and $Z$ preserve the derivative bound and the target value.
We thus obtain
\begin{equation}
\begin{gathered}
 W(\xv_0)>C_*-\varepsilon,\qquad W|_Z=0,\\
 q:=c+f_uW\ge\alpha c.
\end{gathered}
 \label{eq:erev-margin}
\end{equation}
This is a valid, nearly optimal polynomial bound. It remains to
produce a finite SOS identity for its residual $q$.

\emph{2.~Extract the zeros of the residual and obtain a positive matrix.}
Step~1 gives a valid polynomial lower bound, but we still need a finite SOS
representation of its residual $q$. The positivity theorem we invoke to obtain
such a representation requires strict positivity throughout the domain.
Although the margin $q\ge\alpha c$ ensures strict positivity off $F$, the
residual $q=c+f_uW$ vanishes on $F$, where $c=0$ and the derivative bound
forces $f_uW=0$. We therefore factor out these known zeros into amplitude
factors and construct a strictly positive matrix, whose SOS certificate is
obtained in Step~3.

For a one-variable illustration{ with real scalar variable $\tau$}, consider
$q_{\rm toy}(\tau)=\tau^2a(\tau)$ with $a(\tau)>0$.
If we obtain $a(\tau)=\sum_k h_k(\tau)^2$, then multiplication by
$\tau^2$ gives
\[
q_{\rm toy}(\tau)=\sum_k[\tau h_k(\tau)]^2.
\]
For example, $a(\tau)=1+\tau^2$ yields
$q_{\rm toy}(\tau)=\tau^2+(\tau^2)^2$.
Multiplying each square factor by $\tau$ reproduces the required
zero at $\tau=0$. {For the physical problem, let
$r=(\Re r_1,\Im r_1,\Re r_2,\Im r_2)^T$ collect the four real Rydberg-amplitude
components. We use the analogous factorization $q=r^TAr$, with $A$ a real
symmetric $4\times4$ polynomial matrix: the outer factors $r$ account for
the zeros, while we seek a strictly positive $A$ to which a positivity
theorem can be applied.}

{Let $\mathcal I_F$ be the ideal generated by the entries of $r$,
so $F=\{r=0\}$.}
Membership in $\mathcal I_F^3$ means that a polynomial is a sum
of terms containing at least three of these real amplitude factors.
{Recall also the progress observable
$z=\bar g_2g_1^2+\sqrt2\,\bar r_2g_1r_1$ of \eqref{eq:erev-progress}.}
Since $c\in\mathcal I_F^2$ and \eqref{eq:erev-progress} gives
$f_uz\in\mathcal I_F^3$, the definition of $J$ places every
coordinate of $DJf_u$ in $\mathcal I_F^3$. Because $G$ preserves
$\mathcal I_F$, group averaging also gives $f_uW\in\mathcal I_F^3$.
Thus the derivative term contributes only terms of degree at least
three in $r$, whereas the exposure rate is the positive quadratic form
$c=r^TDr$, with $D=\operatorname{diag}(1/2,1/2,1/4,1/4)$.
{As in the stationary program, $\mathcal I$ denotes the ideal
generated by the normalization polynomials
$|g_k|^2+|r_k|^2-1$ ($k=1,2$) and $|u|^2-1$.}
Extracting two amplitude factors from each term gives a symmetric
polynomial matrix $A_0$ such that
$q\equiv r^TA_0r\pmod{\mathcal I}$ and $A_0|_F=D$.
Here equality modulo $\mathcal I$ means equality on the normalized
state--control set. In particular, $A_0$ is positive definite near $F$.

Away from $F$, positivity of the scalar $r^TA_0r$ controls only
the matrix direction $r$ at each point. To obtain positivity in
the other directions without changing $q$, {let $I_4$ denote
the $4\times4$ identity matrix and $\Lambda\ge0$ an adjustable real coefficient,
and} define
\begin{equation}
\begin{gathered}
 A_\Lambda=A_0+\Lambda(\|r\|^2I_4-rr^T),\\
 r^TA_\Lambda r\equiv q\pmod{\mathcal I}.
\end{gathered}
 \label{eq:erev-matrix}
\end{equation}
The added matrix annihilates $r$, which explains why the scalar
residual is unchanged. On a vector perpendicular to $r$, it acts
as multiplication by $\Lambda\|r\|^2$. For $\Lambda\ge0$ this
addition is PSD, so it preserves the positivity
already established near $F$. Outside a sufficiently small
neighborhood of $F$, we have $\|r\|\ge\delta>0$. Moreover,
$q\ge\alpha c\ge(\alpha/4)\|r\|^2$ gives
\[
(r/\|r\|)^TA_\Lambda(r/\|r\|)\ge\alpha/4.
\]
The perpendicular block increases by $\Lambda\|r\|^2I${,
where $I$ is the identity on the three-dimensional subspace perpendicular
to $r$}, while
the radial and cross entries remain unchanged. Those entries
are bounded on the compact domain, so the Schur complement makes
$A_\Lambda$ positive definite throughout the normalized
state--control set for sufficiently large $\Lambda$.

\emph{3.~Convert the positive matrix into an SOS identity for the residual.}
We now have the same residual $q=r^TA_\Lambda r$ on the domain,
with $A_\Lambda$ strictly positive definite there.
{Here $x_{\mathbb R}\in\mathbb R^8$ collects the real and imaginary
parts of the four complex state amplitudes.}
{The normalization equations give
$3-\|x_{\mathbb R}\|^2-|u|^2\equiv0\pmod{\mathcal I}$,
so a constant minus the squared coordinate norm is a sum of squares modulo
$\mathcal I$. This verifies the Archimedean condition required by the matrix
Positivstellensatz~\cite{Scherer06}.}
That theorem supplies polynomial matrices $H_k$ satisfying
\begin{equation}
\begin{gathered}
 A_\Lambda\equiv\sum_k H_k^TH_k\pmod{\mathcal I},\\
 q\equiv\sum_k\|H_kr\|^2\pmod{\mathcal I}.
\end{gathered}
 \label{eq:erev-sos}
\end{equation}
The second identity follows by multiplying the first on the left
by $r^T$ and on the right by $r$: each term becomes
\[
r^TH_k^TH_kr=(H_kr)^T(H_kr).
\]
This is the sum of the squares of the polynomial entries of $H_kr$.
Every such entry vanishes on $F$ because $r=0$ there.
We have therefore obtained an SOS certificate for $c+f_uW$
without altering $W$ or its nearly optimal initial value.
Averaging the Gram representations supplies the symmetry blocks
of Appendix~\ref{app:split_certificate} without increasing degree.
{At hierarchy order $d$, the stationary program allows
$\deg W\le2d-1$ and SOS factors of degree at most $d$.
Since the constructed polynomial and its square factors have finite degree,
some sufficiently large $d$ includes them all, proving feasibility in
\eqref{eq:erev-stationary}.}
\end{proof}

{Recall that $L_d^{\mathrm{unsplit}}$ is the supremum of
$W(\xv_0)$ over feasible order-$d$ stationary certificates.}
Lemma~\ref{lem:erev-polynomial} yields
$L_d^{\mathrm{unsplit}}>C_*-\varepsilon$ for all sufficiently large $d$.
{Increasing $d$ retains every previously feasible certificate
(nesting), while each feasible certificate bounds every successful pulse
from below (weak duality). Hence
$L_d^{\mathrm{unsplit}}\le L_{d+1}^{\mathrm{unsplit}}\le C_*$;
the preceding estimate for arbitrary $\varepsilon>0$ then gives convergence
to $C_*$ and proves Theorem~\ref{thm:asymptotic_completeness}.}

\subsection{Convergence to the minimum exposure on a fixed partition}

We show that partitioning preserves convergence to the
minimum exposure $C_*$ without refining the partition.

\begin{corollary}[Fixed exposure partitions]
Fix any finite partition with positive cell widths and ceiling
$U_0>C_{\mathrm{feas}}$, where $C_{\mathrm{feas}}$ is the cost of a known
feasible pulse. For the partitioned program of Sec.~\ref{sec:exposure_partition} and Appendix~\ref{app:split_certificate},
\begin{equation}
\begin{gathered}
 0\le L_d^{\mathrm{unsplit}}\le L_d^{\mathrm{split}}\le L_{d+1}^{\mathrm{split}}\le C_*,\\
 \lim_{d\to\infty}L_d^{\mathrm{split}}=C_*.
\end{gathered}
 \label{eq:erev-partition}
\end{equation}
\end{corollary}

\begin{proof}
For an arbitrary feasible stationary certificate $W(\xv)$ at order $d$,
set $W_j(\xi,\xv)=W(\xv)$ in each cell $j$, where $\xi\in[0,1]$ is the local
cell coordinate. Because $W_j$ does not depend on $\xi$, the cell operator
satisfies the drift identity $c+{L_j}W_j=c+f_uW$. Interface jumps
$W_{j+1}(0,\xv)-W_j(1,\xv)=0$ vanish identically, and $G$-invariance on the
terminal orbit ensures that the terminal residuals reduce to nonnegative
constants.

On the zero-cost {set} $F=\{r_1=r_2=0\}$, the running cost $c$ vanishes and
control reversal yields $f_{-u}=-f_u$. Feasibility under both $u$ and $-u$
requires $f_uW\ge0$ and $-f_uW\ge0$ on $F\times S^1$, which forces
$f_uW=0$. Consequently, the stationary certificate satisfies
\begin{equation*}
0=c+f_uW=\sum_\ell h_\ell^2\qquad\text{on }F\times S^1.
\end{equation*}
Because each $h_\ell$ is a real polynomial, every factor must vanish on
$F\times S^1$. The {bases obtained by face reduction in}
Appendix~\ref{app:split_certificate} span exactly the subspace of such
vanishing factors at degree at most $d$. Expanding in these bases therefore
preserves every stationary certificate without increasing degree. In the
partitioned program, these identical factors certify each cell when all
interval multipliers are chosen as zero.

Because the initial value
$W_1(0,\xv_0)=W(\xv_0)$ is unchanged, this feasible embedding implies
$L_d^{\mathrm{unsplit}}\le L_d^{\mathrm{split}}$, while degree nesting
immediately yields monotonicity $L_d^{\mathrm{split}}\le
L_{d+1}^{\mathrm{split}}$. Furthermore, because $U_0>C_{\mathrm{feas}}\ge C_*$, we have
$U_0-C_*>0$. Choose any $\varepsilon$ satisfying $0<\varepsilon<U_0-C_*$. Since
$C_*$ is the infimum over actual feasible physical pulses, there exists an
actual feasible physical pulse $u_\varepsilon$ with
$C[u_\varepsilon]<C_*+\varepsilon$; otherwise $C_*+\varepsilon$ would be a
strictly larger lower bound on all physical pulse costs, contradicting the
definition of infimum. Adding $C_*$ to $\varepsilon<U_0-C_*$ gives
$C_*+\varepsilon<U_0$. Thus this chosen pulse satisfies the ceiling:
\begin{equation*}
C[u_\varepsilon]<C_*+\varepsilon<U_0.
\end{equation*}
Restricting the feasible set cannot lower its infimum. To explicitly exclude
raising it, suppose the restricted infimum were $C_*+\eta$ for $\eta>0$.
Choosing $0<\varepsilon<\min\{\eta,U_0-C_*\}$, the construction above then
gives a retained pulse with cost below that proposed infimum, a contradiction.
We conclude that the restricted infimum is $C_*$ and weak duality gives
$L_d^{\mathrm{split}}\le C_*$. Together with stationary completeness
$\lim_{d\to\infty}L_d^{\mathrm{unsplit}}=C_*$
(Theorem~\ref{thm:asymptotic_completeness}), the sandwich
$L_d^{\mathrm{unsplit}}\le L_d^{\mathrm{split}}\le C_*$ establishes
$\lim_{d\to\infty}L_d^{\mathrm{split}}=C_*$.
\end{proof}

\section{Methods and computational reproducibility}
\label{sec:methods}

\begin{sloppypar}
This appendix describes the numerical pulse searches, certificate generation,
and independent verification procedures supporting the exposure bounds in the main text.

To examine whether longer gate durations reduce exposure, we optimize
piecewise-constant phase pulses locally. The sector dynamics of Eq.~\eqref{eq:schrod}
are propagated by closed-form matrix exponentiation on each constant-drive segment.
The duration--exposure landscape across $S\in[7.612,12]$ was mapped via a $96$-segment
multi-start local search, combining $14$ randomized starts of the limited-memory BFGS
(L-BFGS) algorithm followed by sequential least squares programming (SLSQP), with an
additional fine-grid point at $S=7.7$ and a warm-started continuation to $S=16$.
The $192$-segment refinement at duration $S=7.725$ is displayed in Fig.~\ref{fig:pulse}.
{For the $192$-segment pulse, the discrete
Karush--Kuhn--Tucker (KKT) stationarity residual for the segment phases has
maximum magnitude $1.85\times10^{-8}$.}
For the duration--exposure comparison in Sec.~\ref{sec:exposure_dynamics},
the speed-limit candidate has $S\approx7.612$ and $C\approx2.9571$,
while the illustrative pulse has $S=7.725$ and $C\approx2.9360$.
The exposure reduction is $0.71\%$ and the duration increase is $1.48\%$.
In the error model $\Gamma\TR+\gamma T$, the longer pulse is preferable when
\begin{equation}
\frac{\gamma}{\Gamma} < \frac{2.9571 - 2.9360}{7.725 - 7.612} \approx 0.187.
\end{equation}
All serialized waveform arrays, landscape datasets, and figure
generation routines are provided in the companion repository~\cite{repo_pub}.

The unsplit moment-SOS hierarchy was solved numerically through relaxation
order $d=11$, with solver settings given in Table~\ref{tab:sos_progression}. The cumulative-exposure
partitioned programs (Table~\ref{tab:sos_comparison}) used the same solver on the rational
knots~\eqref{eq:split_grid}. The resulting floating-point solutions were projected onto
$\mathbb Q(\sqrt2)$ to produce the order-$6$ five-cell and six-cell
certificates and the order-$7$ five-cell certificate of
Theorem~\ref{thm:split_exact}. The order-$6$ serialized
certificates, accompanied by paired verification logs and a SHA-256 checksum manifest, are
archived in the \texttt{certificates/} directory of the repository~\cite{repo_pub}.

\begin{table*}[t]
\centering
\caption{Detailed progression of the unsplit stationary moment-SOS hierarchy through
relaxation order $11$, where the maximum moment degree at relaxation order $d$ is $2d$.
All SDPs were solved with MOSEK 11.2 (32 threads).
The ``Moments'' column counts the charge-invariant joint state-control monomials through
degree $2d$, retaining one representative per conjugate pair
{after the basis rephasing described in Sec.~\ref{sec:sos_reductions}}.
The ``Liouville'' column counts invariant state-only test monomials of degree up to $2d-1$,
each generating one stationarity constraint through the chain rule.
The reported numerical relaxation values $\widehat{L}_d$ increase with $d$.
These floating-point objectives are not certified lower bounds.
The last column gives $C_{\mathrm{up}}-\widehat{L}_d$, where
$C_{\mathrm{up}}$ is the certified upper bound in Eq.~\eqref{eq:upper_bound}.
\textbf{The ``Blocks'', ``Max Block'', and ``PSD Entries'' columns
report the block structure of the drift Gram matrix $Q_{\mathrm{drift}}$ after
$U(1)\times U(1)\rtimes\mathbb{Z}_2$ symmetry reduction and {face reduction}.}}
\label{tab:sos_progression}
\vspace{0.3em}
\footnotesize
\setlength{\tabcolsep}{3.5pt}
\begin{tabular*}{\textwidth}{@{\extracolsep{\fill}}rcrcrcrr@{}}
\toprule
Order $d$ & Moments & Liouville & Blocks & Max Block & PSD Entries & Numerical
$\widehat{L}_d$ & $C_{\mathrm{up}}-\widehat L_d$ \\
\midrule
6  & 561   & 120   & 113 & 75  & 54,078     & 2.8624595698 & 0.0734156560 \\
7  & 1,054 & 203   & 153 & 131 & 206,088    & 2.8819341861 & 0.0539410397 \\
8  & 1,854 & 326   & 199 & 216 & 685,270    & 2.8889310178 & 0.0469442080 \\
9  & 3,080 & 495   & 251 & 342 & 2,039,903  & 2.8981494707 & 0.0377257551 \\
10 & 4,882 & 724   & 309 & 518 & 5,538,371  & 2.9016460002 & 0.0342292256 \\
11 & 7,454 & 1,028 & 373 & 758 & 13,916,543 & \textbf{2.9047452601} &
\textbf{0.0311299657} \\
\bottomrule
\end{tabular*}
\end{table*}

Exact CZ gate closure requires complete population return $r_1(S)=r_2(S)=0$,
imposing four real conditions, and the conditional phase relation
$\Phi_2-2\Phi_1\equiv\pi\pmod{2\pi}$, imposing a fifth. To establish the certified upper
bound of Sec.~\ref{sec:interval_closure}, we construct a Krawczyk enclosure on a
five-parameter slice of the $12{,}672$-segment pulse family at
$S\approx7.739116$~\cite{Krawczyk69}.
The witness vector $p=(S,\chi_{731},\chi_{6676},\chi_{8661},\chi_{11922})$ varies the active
duration $S$ and four selected segment laser phases $\chi_j$, with all remaining
$12{,}668$ segment phases held fixed. Here $\chi_j$ is the constant drive phase on
segment $j$. The interval Jacobian on this slice is nonsingular. The CZ residuals
on this slice are
\begin{equation}
F(p):=\begin{pmatrix}\operatorname{Re}r_1\\\operatorname{Im}r_1\\\operatorname{Re}r_2\\\operatorname{Im}r_2\\\operatorname{Im}(g_2\bar
g_1^2)\end{pmatrix}=0.
\label{eq:F_map}
\end{equation}
A candidate parameter vector $p_0$ is refined by multi-precision Newton iteration until the residual satisfies $\lvert F(p_0)\rvert < 10^{-135}$.
On an interval box $\mathcal{B}\subset\mathbb{R}^5$ of radius $10^{-80}$ centered at the
multi-precision candidate $p_0$, the Krawczyk operator {is defined by}
\begin{equation}
K(\mathcal{B}):=p_0-RF(p_0)+(I-R[DF(\mathcal{B})])(\mathcal{B}-p_0){.}
\end{equation}
{Here $[DF(\mathcal B)]$ is an interval matrix enclosing the
Jacobian $DF(p)$ for every $p\in\mathcal B$, and $R$ is an approximate inverse
of $DF(p_0)$.} Arb {interval} arithmetic at $500$ bits in \texttt{python-flint}~\cite{Johansson17}
establishes the strict contraction
\begin{equation}
\begin{gathered}
K(\mathcal{B})\subset\operatorname{int}(\mathcal{B}),\\
\|I-R[DF(\mathcal{B})]\|_\infty<3.18\times10^{-75}<1,
\end{gathered}
\end{equation}
proving that $F$ possesses a unique root in $\mathcal{B}$.
The certified condition $\operatorname{Re}(g_2\bar g_1^2)<0$ throughout
$\mathcal{B}$ selects the entangling CZ branch. Evaluating the exposure over $\mathcal{B}$ yields an interval centered at $2.9358752258441048994\ldots$ with radius $5.06\times10^{-79}$, whose upper endpoint lies strictly below $C_{\mathrm{up}}$ and establishes the upper bound of Eq.~\eqref{eq:upper_bound}.

The exact dual auditor is implemented as a standalone Rust tool
(\texttt{minexposure}). It rebuilds
the monomial basis, verifies polynomial identities over $\mathbb Q(\sqrt2)$, and confirms
the positive semidefiniteness of Gram matrix blocks directly from serialized certificates
without invoking an SDP solver. Audits of the order-$6$ and order-$7$ five-cell certificates
are described in Appendices~\ref{app:auditor} and~\ref{app:order7_certificate}.
The \texttt{python-flint} checker for the strengthened lower bound of
Theorem~\ref{thm:slack_strengthened} converts projection polynomials to tensor Bernstein
form on exposure and population boxes. It verifies coefficient inequalities and
per-Gram budgets in exact rational arithmetic, checks
{the reachability exclusions derived from $D$},
and {rejects deliberately corrupted test certificates}.
These checks require no floating-point optimizer.
Detailed verification counts, audit logs, reproduction scripts, and
environment specifications are archived in the companion repository~\cite{repo_pub}.
\end{sloppypar}

\end{document}